\PassOptionsToPackage{colorlinks=true, allcolors=blue}{hyperref}
\documentclass{aa}  

\usepackage{graphicx}
\usepackage{multirow}
\usepackage{subcaption} % For subfigures
\usepackage{bm} % bold math
\usepackage{chemformula} % for chemical reactions and molecules naming
\let\ce\ch % for chemical reactions and molecules naming
\usepackage{xspace} % for correct shortcuts spacing from other words and punctuation marks
\newcommand{\kms}{\,km\,s$^{-1}$\xspace} % kilometres per second
\newcommand{\hrs}{\,hrs\xspace} % hours
\newcommand{\pc}{\,pc\xspace} % parsec
\newcommand{\K}{\,K\xspace} % Kelvin
\newcommand{\mK}{\,mK\xspace} % millikelvin
\newcommand{\GHz}{\,GHz\xspace} % Gigahertz
\newcommand{\MHz}{\,MHz\xspace} % Megahertz
\newcommand{\kHz}{\,kHz\xspace} % Kilohertz
\newcommand{\mm}{\,mm\xspace} % millimeter
\newcommand{\Debye}{\,D\xspace} % Debyes
\usepackage{booktabs} % proof for aligning in table
\usepackage{array} % proof for aligning in table
\usepackage{upgreek} % Greek leters not in italics
\usepackage{orcidlink} % to add orcid links within the authors list
\usepackage[colorlinks=true, allcolors=blue]{hyperref}
\usepackage{longtable}
\usepackage{caption}
\usepackage{subcaption}
\usepackage{arydshln} % dashed hline and cline in tables
\usepackage{dashrule}
\usepackage[varg]{txfonts}
\begin{document}

\title{The Galactic centre G+0.633-0.0604 molecular cloud: A new astrochemical gold mine}
\subtitle{II. Chemical richness}
%\subtitle{II. Chemical Inventory: prebiotic species and second interstellar detections}
%\subtitle{II. Chemical Inventory of Prebiotic sand Rare Interstellar Species}

\author{D. San Andr{\'e}s\inst{1}\fnmsep\inst{2}\fnmsep\thanks{Corresponding author: david.sanandres@cab.inta-csic.es}\orcidlink{0000-0001-7535-4397}
        \and
        V. M. Rivilla\inst{1}\orcidlink{0000-0002-2887-5859}
        \and
        L. Colzi\inst{1}\orcidlink{0000-0001-8064-6394}
        \and
        M. Sanz-Novo\inst{3}\orcidlink{0000-0001-9629-0257}
        \and
        S. Mart{\'i}n\inst{4}\fnmsep\inst{5}\orcidlink{0000-0001-9281-2919}
        \and
        \linebreak
        I. Jim{\'e}nez-Serra\inst{1}\orcidlink{0000-0003-4493-8714}
        \and
        S. Zeng\inst{6}\orcidlink{0000-0003-3721-374X}
        }

\institute{Centro de Astrobiolog{\'i}a (CAB), CSIC-INTA, Carretera de Ajalvir km 4, 28850 Torrej{\'o}n de Ardoz, Madrid, Spain
%\email{david.sanandres@cab.inta-csic.es}
\and
Departamento de F{\'i}sica de la Tierra y Astrof{\'i}sica, Facultad de Ciencias F{\'i}sicas, Universidad Complutense de Madrid, 28040 Madrid, Spain
\and
Center for Astrochemical Studies, Max-Planck-Institut f{\"{u}}r extraterrestrische Physik, Giessenbachstrasse 1, Garching bei Munchen, 85748, Germany 
\and
European Southern Observatory, Alonso de C{\'o}rdova, 3107, Vitacura, Santiago 763-0355, Chile
\and
Joint ALMA Observatory, Alonso de C{\'o}rdova, 3107, Vitacura, Santiago 763-0355, Chile
\and
Star and Planet Formation Laboratory, Pioneering Research Institute (PRI), RIKEN, 2-1 Hirosawa, Wako, Saitama, 351-0198, Japan
}

\date{Received 11 July 2026; Accepted 17 September 2026}

% \abstract{}{}{}{}{}
% 5 {} token are mandatory
 
\abstract
{
In recent years, the inventory of interstellar molecules has rapidly grown, yet detections of complex species are confined to a few chemically rich sources, raising the question of whether such complexity is genuinely rare or shaped by observational biases.
%In recent years, the inventory of interstellar molecules has rapidly grown, yet these discoveries remain confined to a few rich sources, raising the question of whether such chemical complexity is intrinsically rare or shaped by observational biases.
%In recent years, the inventory of interstellar molecules has rapidly grown, yet the number of sources hosting these detections remains very limited, restricting our understanding of interstellar chemical complexity.
%Over the last few years, the inventory of interstellar molecules has rapidly grown. However, the number of sources hosting these new detections remains very limited, driving the search for new chemically-rich sources essential to expand our knowledge about interstellar chemical complexity.
%Astrochemistry is living its golden age, evidenced by the exponential growth over the last few years in the inventory of interstellar molecules. 
%Over the last few years, the inventory of interstellar molecules has rapidly grown. However, the number of sources hosting these new detections is very scarce, driving the search for new chemically-rich sources a necessity to expand our knowledge about interstellar chemical complexity.
}
%on these species interstellar chemistry.}
{
We present the Galactic Centre G+0.633-0.0604 molecular cloud, located in the southern part of the Sgr B2 complex, as a new and very promising laboratory for the discovery of new interstellar molecules.
%We provide a new very prolific source as a highly-promising test-bed for the discovery of new interstellar molecules: the Galactic Centre G+0.633-0.0604 molecular cloud, located in the southern part of the Sgr B2 complex.
%We present a new very promising candidate, the Galactic Centre G+0.633-0.0604 molecular cloud, located in the southern part of the Sgr B2 complex. G+0.633 stands as a powerful \ce{HNCO} emitter, a shock tracer, and exhibits an almost identical chemical feedstock, in both complex and prebiotically relevant molecules, as the well-known Galactic Centre G+0.693-0.027 molecular cloud.
}
{
We analysed data from a high-sensitivity molecular line survey of G+0.633, covering nearly 100\GHz of aggregated bandwidth over the 31$-$275\GHz frequency range with the Yebes 40m, IRAM 30m and APEX radio telescopes.
%We analysed data from a new high-sensitivity molecular line survey of G+0.633, carried out with the Yebes 40m ($\sim$31.1$-$50.4\GHz), IRAM 30m ($\sim$72.0$-$116.6\GHz) and APEX radio telescopes ($\sim$209.9$-$218.3, 226.2$-$234.6, 249.9$-$258.8 and 266.2$-$275.1\GHz).
}
{ 
%stands as a powerful \ce{HNCO} emitter and 
%G+0.633 exhibits an almost identical chemical feedstock, in both complex and prebiotically relevant molecules, as the well-known Galactic Centre G+0.693-0.027 molecular cloud. 
G+0.633 exhibits an exceptionally rich molecular feedstock comparable to that of the chemically rich G+0.693 cloud in the northern part of Sgr B2, establishing it as its ``astrochemical twin''.
We report new robust interstellar detections of 15 astrochemically relevant species, most of prebiotic interest, previously identified only in G+0.693 or a very limited number of sources. These species include urea (\ce{NH2C(O)NH2}), $C$- and $N$-cyanomethanimine ($Z,E$-\ce{HNCHCN}, \ce{H2CNCN}), ethanolamine (\ce{NH2CH2CH2OH}), glycolamide ($sp$-\ce{NH2C(O)CH2OH}), 
%vinylamine (\ce{C2H3NH2}), 
carbonic acid ($sp$-$ap$-\ce{HOCOOH}), dimethyl sulphide (\ce{CH3SCH3}) and thionylimide ($Z$-\ce{HNSO}). 
%The current molecular inventory of G+0.633 now comprises more than 120 species and over 60 isotopologues.
These detections increase the current molecular inventory of G+0.633 to more than 120 species and over 60 isotopologues.
%These add into a total molecular census of $>$120 molecules and $>$60 isotopologues identified in this cloud.
}
{
Our results reveal that the extraordinary molecular complexity observed in G+0.693 is not unique, but may instead reflect a natural outcome of shock-dominated chemistry in the Galactic Centre. The G+0.633 and G+0.693 clouds exhibit remarkably similar excitation conditions and comparable molecular abundances, establishing them as an ideal benchmarking pair for astrochemical studies. Moreover, G+0.633 offers a clear advantage for molecular observations over G+0.693 owing to the significantly narrower linewidths of its molecular line emission, nearly half as broad as those characteristic of G+0.693, making it an exceptionally powerful target for guiding the search for increasingly complex molecules in the ISM.
%Moreover, G+0.633 offers a clear observational advantage over G+0.693 owing to its half narrower linewidths, making it an exceptionally powerful target for guiding the search for increasingly complex molecules in the ISM.
%The comparison of the G+0.633 and G+0.693 molecular clouds indicates that these two sources share very similar excitation conditions and comparable molecular abundances, establishing them as an ideal benchmarking pair for astrochemical studies. Our results further suggest that the extraordinary molecular complexity observed in G+0.693 is not unique, but widespread across the Galactic Centre. Nonetheless, G+0.633 exhibits a clear observational advantage over G+0.693 owing to its half narrower linewidths, making it an exceptionally powerful target for optimising the search for increasingly complex interstellar molecules.
%The comparison of the G+0.633 and G+0.693 molecular clouds indicates that these two sources share very similar excitation conditions and comparable molecular abundances, which make them an ideal benchmarking pair for astrochemical studies. Nonetheless, G+0.633 exhibits a clear observational advantage over G+0.693 given its half narrower linewidths, endowing it with great potential for optimising the search for increasingly complex interstellar molecules.
}


\authorrunning{San Andr{\'e}s et al.}\titlerunning{The chemical wealth of the Galactic centre G+0.633–0.0604 molecular cloud}
%\authorrunning{San Andr{\'e}s et al.}\titlerunning{The chemical wealth of the G+0.633–0.0604 molecular cloud: prebiotic species and second reports in the ISM}
%\authorrunning{San Andr{\'e}s et al.}\titlerunning{The rich inventory of prebiotic and rare molecules of the Galactic Centre G+0.633–0.0604 molecular cloud}
\maketitle
%
%________________________________________________________________

\nolinenumbers

\section{Introduction}

In recent years, astrochemistry has witnessed a remarkable increase in the number of molecules that have been detected in the interstellar medium (ISM), with more than 110 new species reported (out of the $\sim$350 in total\footnote{\label{footnote1}\url{https://cdms.astro.uni-koeln.de/classic/molecules}}) since the latest published record \citep{McGuire2022}. Among them, the identification of complex organic molecules (COMs, denoting carbon-based molecules with at least six atoms; \citealt{Herbst2020}) has particularly soared, constituting nearly half of the current registered molecular census \citep{Jimenez-Serra2025}. To this list, several complex molecules very relevant in the context of prebiotic chemistry 
such as malononitrile (\ce{CH2(CN)2}; \citealt{Agundez2024}), glycolamide ($sp$-\ce{NH2C(O)CH2OH}; \citealt{Rivilla2023}), $n$-propanol ($n$-\ce{C3H7OH}, \citealt{Jimenez-Serra2022}), $N$-cyanomethanimine (\ce{H2CNCN}; \citealt{SanAndres2024}) or ethanolamine (\ce{NH2CH2CH2OH}; \citealt{Rivilla2021_Ethanolamine}) have been recently added, supporting the plausible exogenous-delivery scenario for the origin of life on Earth %\citep{Chyba&Sagan1992, Rubin2019, Vazart2019}. 
\citep{Chyba&Sagan1992, Rubin2019}. Furthermore, the continuous refinement of astronomical instrumentation and improvements in observational sensitivity,
%the continuous refinement of the astronomical instrumentation and the improvement in sensitivity of the observational data, %(e.g, \citealt{Richard2015 Tercero2021}), 
are enabling the detection of increasingly complex molecules, among which 2-methoxyethanol (\ce{CH3OCH2CH2OH}; \citealt{Fried2024}), 2,5-cyclohexadien-1-thione (an isomer of c-\ce{C6H6S}; \citealt{Araki2026}), cyanocoronene (\ce{C24H11CN}; \citealt{Wenzel2025}) and the isomers of cyanoacenaphthylene (\ce{C12H7CN}; \citealt{Cernicharo2024, Cernicharo2026}) and cyanopyrene (\ce{C16H9CN}; \citealt{Wenzel2024a, Wenzel2024b}) stand out.

Nonetheless, despite the rapid growth in the inventory of interstellar molecules, the catalogue of sources hosting these new detections is still very sparse. This imposes severe limitations on the understanding of the chemistry linked to these species under interstellar conditions, 
%driving the search for new chemically-rich sources a prime necessity to shed light on this issue. 
making the search for new chemically-rich sources a key step towards identifying the physical and chemical processes that drive interstellar molecular complexity.
To date, only three sources have emerged as the main molecular repositories with more than 100 identified species:
%To date, there are only three sources at the forefront of most pioneer discoveries with more than 100 species already identified:
%Up to date, three sources are in the lead with more than 100 species already identified: 
the TMC-1 dark cloud (e.g., \citealt{McGuire2020}; \citealt{Cernicharo2024}), the Sgr B2 hot cores (e.g., \citealt{Belloche2013, Belloche2019}), and the Galactic Centre (GC) G+0.693-0.027 molecular cloud (hereafter G+0.693; e.g., \citealt{Jimenez-Serra2020, Rivilla2022_nitriles}). Among these, the G+0.693 cloud, located in the northern part of the Sgr B2 complex ($\sim$55\arcsec\xspace or $\sim$2\pc north-east of the Sgr B2(N) hot core), places itself as a particularly promising source. 
%indeed emerged as one of the richest molecular reservoirs known in the ISM.
%has emerged as one of the most prolific molecular reservoirs known in the ISM. 
Its peculiar combination of 
%high kinetic temperatures, $T_\text{kin}$, of $\sim$70$-$140\K \citep{Zeng2018, Colzi2024}) sharply contrasting with the very low excitation temperatures, $T_\text{ex}$, of $\sim$3$-$20\K (\citealt{Rivilla2020b, Colzi2022}) due to the relatively low \ce{H2} density ($\sim$10$^{4} - 10^{5}$ $\text{cm}^{-3}$; \citealt{Zeng2020, Colzi2024}), 
subthermal excitation (e.g., \citealt{Colzi2022, SanAndres2023})
and widespread low-velocity shocks (\citealt{Martin2008, Zeng2020, Colzi2024}), has established it as a benchmark laboratory for the exploration of molecular diversity in the interstellar medium. To date, more than 150 molecular species have been identified towards G+0.693, including dozens of COMs, rare isotopologues and numerous molecules not yet detected elsewhere in the ISM, with several of high prebiotic relevance
%that comprise many rare isotopologues and up to 40 COMs, many of which have not been reported towards any other source yet 
(e.g., \citealt{Rivilla2019, Rivilla2020b, Rivilla2021_HNCN, Rivilla2021_Ethanolamine, Rivilla2022_ethenediol, Rivilla2022_PO+, Rivilla2023, Zeng2021, Zeng2025, Bizzocchi2020, Rodriguez-Almeida2021_thiols, Rodriguez-Almeida2021_C2H5NCO, Sanz-Novo2023, Sanz-Novo2024_HOCS+, Sanz-Novo2024_HNSO, Sanz-Novo2025_DMS, Sanz-Novo2026_C3H6O2isomers, Fatima2023, Rey-Montejo2024, SanAndres2024, Jimenez-Serra2026}).

To address the current scarcity of chemically-rich interstellar laboratories, in this work, the second of a series, we expand on the chemical characterisation of the recently identified G+0.633-0.0604 molecular cloud (hereafter G+0.633). Located $\sim$4\arcmin\xspace ($\sim$10\pc) south of G+0.693, at the southernmost edge of the Sgr B2 complex within the Sgr B2(DS) region, G+0.633 offers a unique opportunity to investigate whether the extraordinary chemistry observed in G+0.693 is an isolated case or instead a natural outcome within shock-dominated environments such as the Sgr B2 complex. In the first article of this series (\citealt{SanAndres2026_G0633-PaperI}; hereafter Paper I), we investigated the physical properties of G+0.633, identifying three velocity components along the line of sight. The dominant component, centred at $v_{\rm LSR}\sim48.5$\kms and displaying relatively narrow linewidths ($\sim$10\kms), is characterised by very bright \ce{HNCO} emission, indicative of a shock-triggered nature, and by physical properties remarkably similar to those measured in G+0.693. These results suggest that both clouds may represent analogous evolutionary stages within the large-scale cloud–cloud collision scenario proposed for the Sgr B2 complex \citep{Colzi2024}, making G+0.633 a likely southern counterpart of G+0.693.

Building on this picture, the molecular line survey of G+0.633 has revealed a remarkably rich chemistry, with a diverse census comprising more than 120 molecules and nearly half as many isotopologues. These observations have already uncovered notable molecules in this source, including aromatic species such as benzonitrile (c-\ce{C6H5CN}; \citealt{Rivilla2026_benzo}), highlighting a level of molecular complexity comparable to that of G+0.693. Importantly, the survey provides second interstellar detections of several species previously known only towards G+0.693, as well as new detections of molecules reported in just a handful of sources. These molecules constitute the main focus of this work. In particular, we present the analysis of 15 species of high prebiotic and astrobiological interest, including urea (\ce{NH2C(O)NH2}), ethanolamine (\ce{NH2CH2CH2OH}), glycolamide ($sp$-\ce{NH2C(O)CH2OH}) and dimethyl sulphide (DMS, \ce{CH3SCH3}). Together, these findings establish G+0.633 as an ``astrochemical twin'' of G+0.693 within the ISM of the GC.

This paper is organised as follows. In Sect.~\ref{sec:observations}, we present the observational data of the molecular line survey conducted towards G+0.633. In Sect.~\ref{sec:analysis_and_results}, we describe its underlying analysis and provide the main results. In Sect.~\ref{sec:discussion}, we compare the chemical inventories of G+0.633 and G+0.693, and contrast the physical properties of the observed molecular emission. Finally, in Sect.~\ref{sec:summary_and_conclusions}, we outline our conclusions.

%% ! Esto para las conclusiones u otro punto del paper
%On top of that, this cloud appears as an excellent candidate for optimising the search for new complex molecules, even surpassing the potential of G+0.693. This is because mainly due to its narrower linewidths ($\sim$10\kms) in comparison to those found in G+0.693 ($\sim$18$-$25\kms; \citealt{Rivilla2020b, Colzi2022}), which greatly reduce line blending. 

%__________________________________________________________________

\section{Observations}
\label{sec:observations}

We have analysed data from the newly conducted ultradeep spectral line survey of G+0.633, which combines single-dish observations from the Yebes 40m (Guadalajara, Spain), IRAM\footnote{Institut de radioastronomie millimétrique} 30m (Granada, Spain) and APEX\footnote{Atacama Pathfinder EXperiment} (Chajnantor, Chile) radio telescopes. In all cases, we conducted position-switching observations centred at $\alpha_\text{ICRS} = 17^\text{h}47^\text{m}21.18^\text{s}$, $\delta_\text{ICRS} = -28^\text{o}25'36.99''$, with the off position shifted by $\Delta\alpha = -874''$, $\Delta\delta = 540''$ (i.e., $\alpha_\text{ICRS}(\text{OFF}) = 17^\text{h}46^\text{m}14.92^\text{s}$, $\delta_\text{ICRS}(\text{OFF}) = -28^\text{o}16'36.99''$). 
%This particular reference position is the same previously adopted for the neighbouring G+0.693 cloud surveys (e.g., \citealt{Rivilla2021_Ethanolamine, Sanz-Novo2023}), devoid from emission of abundant molecules such as \ce{CS} and \ce{HC3N}.
The line intensity of the spectra was directly measured in antenna temperature ($T_\text{A}^*$) units, since the molecular emission towards G+0.633 is extended compared to the primary beam of the three telescopes (Paper I).
%\citealt{Brunken2010, Jones2012, Li2020, Santa-Maria2021, Colzi2024}). 
In the following, we provide a summary of the observational survey, with a graphical overview on both the spectral coverage and sensitivity achieved given in Appendix~\ref{appendix:rms}. For further details on both the observations and data reduction process, we refer to Paper I.

The observations on G+0.633 with the Yebes 40m radio telescope (project 24A010; PI: San Andr{\'e}s) were performed between February and April 2024, accumulating $\sim$28\hrs on-source. The Nanocosmos Q band (7\mm) HEMT receiver was used to cover the entire Q band (31.07$-$50.42\GHz), providing a raw frequency resolution of 38\kHz \citep{Tercero2021}. The spectra were later smoothed to a resolution of 256\kHz ($\sim$1.54$-$2.56\kms in the range observed), which is enough to properly resolve the observed linewidths of $\gtrsim$10\kms (Paper I). We achieved rms noise levels between 0.5$-$1.3\mK across the $\sim$31$-$48\GHz range, rising to $\sim$2$-$2.6\mK in the Q-band upper end ($>$48\GHz) at this resolution (see Appendix~\ref{appendix:rms}). The half power beam width (HPBW) of the Yebes 40m telescope varies from $\sim$55$''$ at 31\GHz down to $\sim$35$''$ at 50\GHz.

The IRAM 30m observations of G+0.633 were conducted under projects 058-21 (PI: Rivilla) and 155-24 and 055-25 (PI: San Andr{\'e}s), aimed at covering the entire 3\mm band ($\sim$72$-$116.6\GHz) with very high sensitivity. The observations were gathered in September 2021 (project 058-21), and between April 2025 and November 2025 (projects 155-24 and 055-25), for a total time on-source of $\sim$57.5\hrs. In all cases, we used the broadband mm-wave heterodyne Eight MIxer Receiver (EMIR; \citealt{Carter2012}) in combination with the Fast Fourier Transform Spectrometer FTS200, which provided a raw channel width of $\sim$195\kHz. 
%The observations on project 058-21 were gathered during a three-night observational campaign on September 2021, covering the lower part of the E0 (3\mm) band ($\sim$72$-$103\GHz) with a total time on source of $\sim$10.5\hrs. The observations on project 155-24 were conducted along eight days between April (5 runs) and June 2025 (2 runs), and covered the upper part of the E0 band ($\sim$84$-$116.6\GHz range) with a total time on source of $\sim$10.5\hrs. Finally, project 055-25 was observed for 20 days (15 in June 2025, 5 in November 2025) achieving a total time on source of $\sim$36.5\hrs, and traced the entire 3\mm band to homogenise the sensitivity reached across it. 
We smoothed the reduced spectra to 676\kHz, which translates into velocity resolutions of 1.74$-$2.82\kms across the covered range. The final rms of the spectra varies $\sim$0.4$-$1.2\mK in the $\sim$72$-$114\GHz range, and rises up to $\sim$6.2\mK at the high end of the 3\mm band ($>$114\GHz) due to increased atmospheric opacity (see Appendix \ref{appendix:rms}). The HPBW of the IRAM 30m telescope ranges from $\sim$34$''$ down to $\sim$21$''$ along the observed frequency range.

The APEX observations of G+0.633 (project ID O-0109.F-9309A-2022; PI: Rivilla) were taken between April and October 2022, for a total time on-source of $\sim$4.5\hrs. We made use of the 230\GHz-band New First Light APEX Submillimeter Heterodyne (NFLASH230\footnote{The nFLASH receiver was delivered by MPIfR Sub-mm technology division: \url{https://www.mpifr-bonn.mpg.de/5278273/nflash}}) receiver connected to two FFTS backends, covering four spectral ranges with a raw spectral resolution of $\sim$250\kHz: 209.92$-$218.33, 226.18$-$234.58, 249.92$-$258.83 and 266.17–275.08\GHz. The spectra were later smoothed to 1.0\MHz, which translates into velocity resolutions of 1.1$-$1.5\kms in the spectral range covered. We achieved a very low rms noise level ranging 0.7$-$1.0\mK along the observed windows at this spectral resolution (reaching a maximum of $\sim$1.5$-$2.1\mK at their borders; see Appendix~\ref{appendix:rms}). The HPBW of these APEX observations is $\sim$22$-$29$''$ in the range traced.

%__________________________________________________________________

\section{Analysis and results}
\label{sec:analysis_and_results}

%On the basis of its peculiar physical properties described in Paper I, 
%The G+0.633 cloud emerges as an outstanding new interstellar laboratory for astrochemistry.
%to search for new complex molecules. 
%Within the current observational survey, 
%Thus far, it gathers the detection of about 120 different molecular species (half of which are COMs) and 
The molecular line survey of G+0.633 reveals an exceptionally rich chemical inventory, comprising about 120 species (roughly half of which are COMs) in addition to nearly half as many associated isotopologues.
%nearly half a hundred 
%nearly half as many of their associated isotopologues. 
While many of these molecules are widespread in the ISM, such as \ce{CH3CN}, \ce{CH3CCH}, \ce{HC3N} or \ce{HNCO} previously analysed in Paper I, the exquisite sensitivity of the observational data has uncovered some species hitherto only identified in a very limited number of interstellar sources. In this work, we focus on 15 of these elusive species (see Table~\ref{tab:detected_molecules_fitting_parameters}), many of high prebiotic relevance, which provide an excellent indicator of G+0.633 astrochemical and astrobiological potential. We note that, given their lower abundances and therefore weaker spectral lines, only the main velocity component from the three characterised in Paper I (i.e., C1) can be clearly discerned, thus being the only one considered in the LTE analysis of these species. Accordingly, molecular abundances relative to \ce{H2} are computed using $N_{\ce{H2}} = (6.0 \pm 1.2)\times10^{22} \; \text{cm}^{-2}$ (see Paper I for C1). The spectroscopic information of all these molecules and their analysed transitions is given as supplementary material in  %Table~\ref{tab:molecules_spectroscopic_info} within 
%Appendix~\ref{appendix:molecular_spectroscopy}
\href{https://zenodo.org/records/20745575?token=eyJhbGciOiJIUzUxMiJ9.eyJpZCI6ImM1NzQyYmY3LWUzODEtNGFiNC04NDE1LTM5NjdiYzViMzY1NSIsImRhdGEiOnt9LCJyYW5kb20iOiIwOTgwYjI1MDAxMDAwZjM4NzNkNzUwM2U3NGZjMzJiYyJ9.tUjjNIOROJge6sYJGloh1mCU7NaM0ebR5RihkQXXHSc0Ta4284wiSYh-Cw1KfKnzWtqjtRA-ddbS4afNHV8hXg}{Zenodo}.

We performed the identification and spectral modelling for all molecular species presented in this work by using the 11 June 2026 version of the Spectral Line Identification and Modelling (SLIM) tool within the \textsc{MADCUBA}\footnote{Madrid Data Cube Analysis (MADCUBA) on ImageJ is a software developed at the Centro de Astrobiología (CAB) in Madrid: \url{https://cab.inta-csic.es/madcuba/}} package \citep{Martin2019}. SLIM collected the available spectroscopic data from the catalogues as the Cologne Database for Molecular Spectroscopy (CDMS, \citealt{Endres2016}), Jet Propulsion Laboratory (JPL, \citealt{Pickett1998}) and the Lille Spectroscopic Database (LSD, \citealt{Motiyenko&Margules2025}) or, in their absence, from the spectroscopic literature; and generates a synthetic spectrum under the assumption of local thermodynamic equilibrium (LTE) that best explains the observed one, while taking into account line opacity. 
%We have not considered hyperfine structure for any molecule, since the broad linewidths observed ($\gtrsim$10\kms) do not allow for it to be resolved in any case. 
To derive the physical parameters describing the molecular emission, we used the automatic fitting routine SLIM-AUTOFIT, which provides the best non-linear least-squares LTE fit to the data using the Levenberg-Marquardt algorithm. 
%The survey presented here is the same dataset analysed in Paper I (which already included all of the species so far identified in the cloud), being line blending systematically, simultaneously and retrogradely assessed as the new molecules were being identified, fitted, and incorporated into it. 
The free parameters are the total column density of the molecule ($N$), the excitation temperature ($T_\text{ex}$), the local standard of rest velocity ($v_\text{LSR}$) and the full width at half maximum ($\text{FWHM}$). We note that the survey presented here is the exact same dataset analysed in Paper I, which thus considers all of the species thus far identified and modelled in the cloud to check for line blending. 
%The results of each species analysis are summarised in Tables~\ref{tab:CO_isotopologues}, \ref{tab:physical_parameters_molecules_modelling} and \ref{tab:detected_molecules_fitting_parameters}.

The LTE analysis of each molecule
%presented in this work 
was done by fitting all previously detected transitions in G+0.693 that are identified with a signal-to-noise ratio (S/N) in integrated intensity $\gtrsim$5 towards G+0.633. For some species, we also incorporated lines that, either unavailable in G+0.693 survey when published or excluded in the fitting because of significant contamination, are now clearly detected in G+0.633 given its much narrower linewidths (see Sect.~\ref{sec:analysis_and_results}).
%and appear unblended (or at most only slightly blended). 
Furthermore, since for most molecules \textsc{AUTOFIT} fails to converge reliably unless the $\text{FWHM}$ is fixed, we followed a similar two-step approach as presented in \citet{SanAndres2024} for G+0.693. For a given molecule, we fitted in a first step, leaving free all parameters, a subset of the most intense and unblended lines (delineated with a $\blacktriangledown$ symbol in the figures)
%lines among all those intended to be used (the most intense and unblended, delineated with a $\blacktriangledown$ symbol in all figures),
constraining the $\text{FWHM}$. Then, in a second step, we repeated \textsc{AUTOFIT} keeping the $\text{FWHM}$ fixed to this value, and using the whole set of identified transitions. We obtain a consistent $\text{FWHM}$ of $\sim$10\kms throughout the survey for the main C1 component (Paper I). Therefore, in those cases for which the two-step analysis is unfeasible, owing to the lack of intense unblended lines,
a $\text{FWHM}$ of 10\kms is assumed directly, and no transitions are marked with the $\blacktriangledown$ symbol in the corresponding figure. The results of the LTE fittings are also given in Table~\ref{tab:detected_molecules_fitting_parameters}.

%\subsection{The rich chemical content of G+0.633}

%\setlength{\tabcolsep}{4.15pt}

\setlength{\tabcolsep}{2.5pt} % changes column separation

\renewcommand{\arraystretch}{1.25} % changes row separation

\begin{table*}[ht!]
 \caption[]{\label{tab:detected_molecules_fitting_parameters}Derived physical parameters from the best LTE fit for the 
 species analysed in this work and comparison to G+0.693.
 %, sorted by increasing complexity.
 %elusive prebiotic and complex interstellar molecules identified in G+0.633, displayed in increasing complexity order.
 }
 \centering
 \begin{tabular}{cccccccccc}
     \hline \hline
     Molecule\tablefootmark{(a), (b)} & Molecule name\tablefootmark{(b)} & $N$ & $T_\text{ex}$ & $\text{FWHM}$ & $v_{\text{LSR}}$ & $N/N_{\ce{H2}}$\tablefootmark{(c)} & $(N/N_{\ce{H2}})_{\, \text{G+0.693}}$ & Ref. \\
     & & $(\times 10^{12} \, \text{cm}^{-2})$ & $(\text{K})$ & $(\text{km} \, \text{s}^{-1})$ & $(\text{km} \, \text{s}^{-1})$ & $(\times 10^{-11})$ & $(\times 10^{-11})$ & \\ 
     \hline
     \multicolumn{8}{c}{ RNA-world precursors } \\
     \hline
     \ce{NH2C(O)NH2} $^\text{(3rd)}$ & urea & 1.7$\pm$0.1 & 9.1$\pm$0.6 & 10.0 & 47.8$\pm$0.3 & 2.8$\pm$0.6 & 5.2$\pm$0.5 & (1) \\
     $Z$-\ce{HNCHCN} $^\text{(2nd)}$ & $Z$-$C$-cyanomethanimine & 20.8$\pm$0.9 & 11.5$\pm$0.6 & 9.9$\pm$0.8 & 48.8$\pm$0.2 & 35$\pm$7 & 55$\pm$9 & (2) \\
     $E$-\ce{HNCHCN} $^\text{(3rd)}$ & $E$-$C$-cyanomethanimine & 4.72$\pm$0.23 & 9.6$\pm$0.5 & 10.2$\pm$0.9 & 48.8$\pm$0.3 & 7.9$\pm$1.6 & 12$\pm$2 & (2) \\
     \ce{H2CNCN} $^\text{(2nd)}$ & $N$-cyanomethanimine & 0.95$\pm$0.08 & 7.6$\pm$0.7 & 10.0 & 47.5$\pm$0.5 & 1.58$\pm$0.34 & 2.1$\pm$0.3 & (2) \\
     \ce{HOCH2CN} $^\text{(5th)}$ & glycolonitrile & 7.4$\pm$0.3 & 13.3$\pm$0.9 & 11.2$\pm$1.1 & 49.0$\pm$0.3 & 12.3$\pm$2.5 & 5.9$\pm$1.7 & (3) \\
     $Z$-\ce{(CHOH)2} $^\text{(2nd)}$ & $Z$-$1{,}2$-ethenediol & 9.9$\pm$0.3 & 13.0$\pm$0.8 & 10.3$\pm$1.2 & 49.0$\pm$0.2 & 16.5$\pm$3.3 & 13$\pm$2 & (4) \\
     \hline
     \multicolumn{8}{c}{ Lipid and protein precursors } \\
     \hline
     \ce{NH2CH2CH2OH} $^\text{(2nd)}$ & ethanolamine & 4.56$\pm$0.17 & 13.2$\pm$0.8 & 10.0$\pm$0.8 & 47.4$\pm$0.2 & 7.6$\pm$1.5 & 11.2$\pm$1.8 & (5) \\
     $sp$-\ce{NH2C(O)CH2OH} $^\text{(2nd)}$ & $syn$-glycolamide & 2.50$\pm$0.16 & 6.0$\pm$0.8 & 10.0 & 48.7$\pm$0.3 & 4.2$\pm$0.9 & 5.5$\pm$1.0 & (6) \\
     \ce{C2H3NH2} $^\text{(2nd)}$ & vinylamine & 12$\pm$1 & 21.2$\pm$1.5 & 10.0 & 50.0$\pm$0.3 & 20$\pm$4 & 33$\pm$4 & (7) \\
     $sp$-$ap$-\ce{HOCOOH} $^\text{(2nd)}$ & $cis$-$trans$-carbonic acid & 2.70$\pm$0.23 & 8.1$\pm$0.9 & 10.0 & 49.1$\pm$0.5 & 4$\pm$1 & 4.7$\pm$0.3 & (8) \\
     \ce{C2H5NCO} $^\text{(2nd)}$ & ethyl isocyanate & 6.2$\pm$0.9 & 24.2$\pm$3.6 & 10.0 & 48.0$\pm$0.9 & 10.3$\pm$2.6 & 6.0$\pm$1.3 & (9) \\
     \ce{(CH3)2CCH2} $^\text{(2nd)}$ & isobutene & 67$\pm$6 & 11.3$\pm$1.8 & 10.0 & 47.2$\pm$0.4 & 112$\pm$24 & 190$\pm$40 & (10) \\
     \hline
     \multicolumn{8}{c}{ \ce{S}-bearing species } \\
     \hline
     \ce{CH3SCH3} $^\text{(2nd)}$ & dimethyl sulfide & 10.3$\pm$0.6 & 7.4$\pm$0.6 & 10.0 & 49.9$\pm$0.4 & 17$\pm$4 & 19$\pm$4 & (11) \\
     $Z$-\ce{HNSO} $^\text{(2nd)}$ & $cis$-thionylimide & 21$\pm$1 & 12.1$\pm$1.3 & 11.2$\pm$0.9 & 49.2$\pm$0.3 & 35$\pm$7 & 60$\pm$10 & (12) \\
     \multirow{2}{*}{\ce{HOCS+} $^\text{(2nd)}$} & carbonyl sulfide & \multirow{2}{*}{2.08$\pm$0.15} & \multirow{2}{*}{28.0} & \multirow{2}{*}{10.0} & \multirow{2}{*}{48.5} & \multirow{2}{*}{3.5$\pm$0.7} & \multirow{2}{*}{7$\pm$2} & \multirow{2}{*}{(13)} \\[-0.75ex]
     & (\ce{O}-protonated) & & & & & & & \\
     \multirow{2}{*}{\ce{HNC^{34}S} $^\text{(2nd)}$} & isothiocyanic acid & \multirow{2}{*}{0.59$\pm$0.12} & \multirow{2}{*}{20.4} & \multirow{2}{*}{10.0} & \multirow{2}{*}{48.5} & \multirow{2}{*}{1.0$\pm$0.3} & \multirow{2}{*}{2.30$\pm$0.34} & \multirow{2}{*}{(14)} \\[-0.75ex]
     & (\ce{^{34}S} isotopologue) & & & & & & & \\
     \hline
 \end{tabular}
  \tablefoot{For comparison, the last two columns provide the molecular abundances derived for each of these species in G+0.693 and the corresponding references, respectively.
  %For comparison, the last column gathers the molecular abundances derived for each of these species in G+0.693, with the corresponding references provided below.
  %following the table order.
  \tablefoottext{a}{Superindexes indicate whether the detection of each species towards G+0.633 is the second (2nd), third (3rd) or $n$th unambiguous in the ISM to the best of our knowledge. Beyond the identifications in G+0.693, additional clear detections include $E$-\ce{HNCHCN} in Sgr B2(N) \citep{Zaleski2013}, \ce{HOCH2CN} towards IRAS16293-2422 B, Serpens SMM1-a and the V883 Ori protoplanetary disc (\citealt{Zeng2019, Ligterink2021, Fadul2025}, respectively), and \ce{NH2C(O)NH2} in Sgr B2(N) \citep{Belloche2019}. 
  %We omitted $sp$-\ce{NH2C(O)CH2OH} recent tentative detection in G358.93 MM1 hot core \citep{Duan2026_glycolamide} since we consider it rather uncertain.
  We omitted $sp$-\ce{NH2C(O)CH2OH} recent report towards G358.93 MM1 hot core \citep{Duan2026_glycolamide} since it is a tentative detection.
  }
  \tablefoottext{b}{Isomers are denoted using the nomenclature recommended in \citet{Rivilla2026_stereoisomers} (prefixes in ``Molecule'' column); however, we also specify their common name in the literature for direct association (prefixes in ``Molecule name'' column).}
  \tablefoottext{c}{Molecular abundances were computed using $N_{\ce{H2}} = (6.0 \pm 1.2)\times10^{22} \; \text{cm}^{-2}$ (Paper I).}
  }
  \tablebib{
  (1)~\citet{Zeng2023}; 
  (2)~\citet{SanAndres2024};
  (3)~\citet{Rivilla2022_nitriles};
  (4)~\citet{Rivilla2022_ethenediol};
  (5)~\citet{Rivilla2021_Ethanolamine};
  (6)~\citet{Rivilla2023};
  (7)~\citet{Zeng2021};
  (8)~\citet{Sanz-Novo2023};
  (9)~\citet{Rodriguez-Almeida2021_C2H5NCO};
  (10)~\citet{Fatima2023};
  (11)~\citet{Sanz-Novo2025_DMS};
  (12)~\citet{Sanz-Novo2024_HNSO};
  (13)~\citet{Lattanzi2024};
  (14)~\citet{Sanz-Novo2024_HOCS+}.
  }
\end{table*}

%\subsubsection{On the foundations of the RNA-world hypothesis: urea, cyanomethanimines, glycolonitrile and $1{,}2$-ethenediol}

\subsection{RNA-world precursors: urea, cyanomethanimines, glycolonitrile and $Z$-$1{,}2$-ethenediol}

The question of how life originated on Earth remains latent in astrobiology. One of the prevailing hypotheses is the so-called RNA-world \citep{Gilbert1986}, which points to this macromolecule as the main
trigger for primordial biological activity. Although the precise chemical pathways that led to the first RNA molecules remain a mystery, various prebiotic chemistry experiments have demonstrated that RNA building blocks, the ribonucleotides, can be synthesized from much simpler molecules 
(\citealt{Powner2009, Patel2015, Becker2019}),
%(\citealt{Powner2009, Patel2015, Becker2016, Becker2019}), 
some of which have already been reported in space, thus supporting the plausible role of interstellar chemistry in the origin of life.

\begin{figure*}[ht!]
    \centering
    \includegraphics[width=0.98\textwidth]{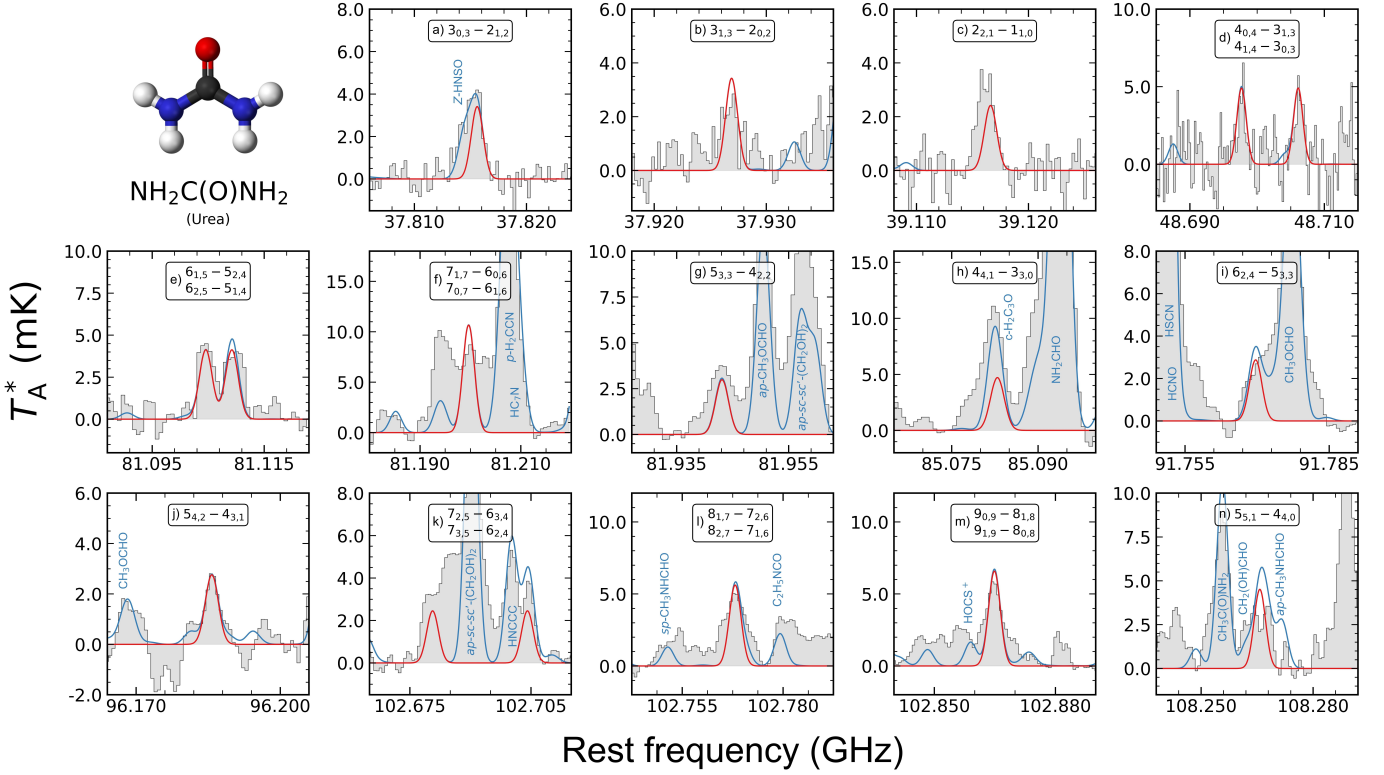}
    \caption{Lines of \ce{NH2C(O)NH2} (urea) detected in G+0.633. The black and grey-shaded histogram indicate the observed spectrum, while the red and blue solid lines represent the best LTE fit obtained for \ce{NH2C(O)NH2}, and the emission from all species already identified in this cloud (blue labels), respectively. 
    %Panel labels indicate the specific rotational transition(s) being plotted.
    %using the common notation for asymmetric tops: $J_{K_\text{a}K_\text{c}}$, with $J$ the total angular momentum of the molecule and $K_\text{a,c}$ its projections along the $a$ and $c$ principal axes. 
   The transitions shown here (specified in the panel labels) were used in the LTE line fitting of this molecule. In this and the following figures, $\blacktriangledown$ symbols denote transitions used in the two-step fitting procedure to constrain the $\text{FWHM}$ in the first step; if absent (as in this case), the fit was performed in a single step adopting $\text{FWHM}=10$\kms directly.
   %but none is marked with a $\blacktriangledown$ symbol since it was directly performed in a single step (adopting $\text{FWHM}=10$\kms). 
    %Their spectroscopic information is given in Table~\ref{tab:urea_rotational_spectroscopy} within Appendix~\ref{appendix:molecular_spectroscopy}. 
    The molecular structure of \ce{NH2C(O)NH2} is shown in the upper left corner (\ce{C} in grey, \ce{N} in blue, \ce{O} in red and \ce{H} in white).
    %, for which we followed the two-steps methodology and being those marked with $\blacktriangledown$ symbols employed to constrain the $\text{FWHM}$ in the first step. Their spectroscopic information is given in Table~\ref{tab:cis-HNSO_rotational_spectroscopy} within Appendix~\ref{appendix:molecular_spectroscopy}. The molecular structure of $Z$-\ce{HNSO} is depicted in the upper left corner (nitrogen atoms in blue, sulphur atoms in yellow, oxygen atoms in red and hydrogen atoms in white colours).
    } 
    \label{fig:urea_detection} 
\end{figure*}

Urea (\ce{NH2C(O)NH2}) is among the most prominent ones, recognized for its essential biological role in nitrogen metabolism, osmoregulation, and waste disposal of present-day animal species. Besides, urea is considered an important molecule in prebiotic chemistry, as it can participate in reaction networks leading to biologically relevant compounds, including pyrimidine nucleobases such as cytosine (\ce{C4H5N3O}) and uracil (\ce{C4H4N2O2}), as well as other RNA-related building blocks (e.g., \citealt{Robertson&Miller1995}; \citealt{Menor-Salvan2009,Menor-Salvan2020}).
%Besides, urea is believed to play a major prebiotic role as a potential precursor of cytosine (\ce{C4H5N3O}) and uracil (\ce{C4H4N2O2}), two of the essential pyrimidine nucleobases that constitute RNA chains (e.g., \citealt{Robertson&Miller1995}, \citealt{Choe2020}). 
To date, urea has only been reported in two astronomical sources, the Sgr B2(N) hot core \citep{Belloche2019} and the G+0.693 molecular cloud \citep{Jimenez-Serra2020, Zeng2023}. Here, we consolidate its third identification in the ISM towards G+0.633 through the detection of 14 transitions covering a broad range of energy levels (with $2 \leq J_\text{up} \leq 9$), as shown in Fig.~\ref{fig:urea_detection}.
%including for the first time several lines at 7\mm that were not targeted in the two previous reports (based on observations at 3\mm). 
We performed the LTE analysis of urea by fitting all these transitions in a single step, directly fixing the $\text{FWHM}$ to 10\kms since most of them are either blended or too weak for the two-step method to succeed, obtaining $N = (1.7 \pm 0.1)\times10^{12}\, \text{cm}^{-2}$, $T_\text{ex} = 9.1 \pm 0.6$\K and $v_\text{LSR} = 47.8 \pm 0.3$\kms. 
%For this species, the  was directly fixed to 10\kms, since most of its detected lines are either blended or too weak for the two-step method to succeed. %to allow for convergence, in accordance with other molecules analysis. 
The resulting abundance of urea relative to \ce{H2} is $(2.8 \pm 0.6)\times10^{-11}$.

Another species of high relevance in the context of the RNA-world hypothesis are cyanomethanimines (\ce{C2H2N2} isomers), which are believed to be potential precursors of adenine (\ce{C5H5N5}; \citealt{Chakrabarti&Chakrabarti2000, Jung&Choe2013}), one of the purine nucleobases present both in RNA and DNA ribonucleotides. Cyanomethanimines emerge as the most stable \ce{HCN}-dimers, and occur in two different structural arrangements: $C$-cyanomethanimine (\ce{HNCHCN}), which in turn can be found in the two distinct $Z$- and $E$- stereoisomeric configurations about the \ce{N=C} bond; and $N$-cyanomethanimine (\ce{H2CNCN}). $Z$-\ce{HNCHCN} is the most energetically stable of them, followed by $E$-\ce{HNCHCN} (with a relative energy of $\Delta E = 309\pm72$\K; \citealt{Takano1990}) and \ce{H2CNCN} ($\Delta E = 3570\pm130$\K; \citealt{Puzzarini2015}). Despite their relative simplicity, all these species have only been reported thus far towards G+0.693 \citep{Rivilla2019, SanAndres2024}, with the exception of $E$-\ce{HNCHCN} which was also identified in the Sgr B2(N) hot core \citep{Zaleski2013}. Here, we show in Figs.~\ref{fig:Z-HNCHCN_detection}, \ref{fig:E-HNCHCN_detection} and \ref{fig:H2CNCN_detection} within Appendix~\ref{appendix:molecules_additional_detection_figures} the newly achieved detection in the ISM of these three isomers ($Z,E$-\ce{HNCHCN} and \ce{H2CNCN}, respectively) towards G+0.633. For all of them, we have clearly identified an almost identical subset of transitions as those targeted in G+0.693 by \citet{Rivilla2019} and \citet{SanAndres2024}, corresponding to $R$-branch $a$-type rotational transitions spanning energy levels with $J_\text{up} = 4-12$ for $Z,E$-\ce{HNCHCN} (including the entire progression of $b$-type transitions with $2 \leq J_\text{up} \leq 6$ for $E$-\ce{HNCHCN}), and $J_\text{up} = 3-9$ for \ce{H2CNCN}. 
%To perform the LTE line fitting for $Z$- and $E$-\ce{HNCHCN}, we followed a similar two-step approach as done in \citet{SanAndres2024}. First, we fitted the most intense unblended lines (shown in panels f), h), i) and m) of Fig.~\ref{fig:Z-HNCHCN_detection} for $Z$-\ce{HNCHCN}; and panels d-f), h) and j) of Fig.~\ref{fig:E-HNCHCN_detection} for $E$-\ce{HNCHCN}) to better constrain these two molecules lines width, obtaining very similar values of $\text{FWHM} = 9.9\pm0.8$\kms and $\text{FWHM} = 10.2\pm0.9$\kms for each of them, respectively. Then, in a second step, we repeated the fit but including the rest of transitions shown in Figs.~\ref{fig:Z-HNCHCN_detection} and \ref{fig:E-HNCHCN_detection}, while keeping fixed the $\text{FWHM}$ to the former values, respectively. 
We performed the LTE analysis of $Z$- and $E$-\ce{HNCHCN} following the two-step approach, constraining their $\text{FWHM}$ by first only fitting the transitions in panels h), i) and l) of Fig.~\ref{fig:Z-HNCHCN_detection} for $Z$-\ce{HNCHCN}, and those in panels e), j), p) and r)-t) of Fig.~\ref{fig:E-HNCHCN_detection} for $E$-\ce{HNCHCN}. We obtained values consistent within uncertainties of $\text{FWHM} = 9.9\pm0.8$\kms and $\text{FWHM} = 10.2\pm0.9$\kms for each of them, respectively. Then, keeping the $\text{FWHM}$ fixed to these values, we derived in the second step (using all transitions shown in Figs.~\ref{fig:Z-HNCHCN_detection} and \ref{fig:E-HNCHCN_detection}): $N = (20.8 \pm 0.9) \times 10^{12}\, \text{cm}^{-2}$, $T_\text{ex} = 11.5 \pm 0.6$\K and $v_\text{LSR} = 48.8 \pm 0.2$\kms for $Z$-\ce{HNCHCN}, whereas $N = (4.72 \pm 0.23) \times 10^{12}\, \text{cm}^{-2}$, $T_\text{ex} = 9.6 \pm 0.5$\K and $v_\text{LSR} = 48.8 \pm 0.3$\kms for $E$-\ce{HNCHCN}. Regarding \ce{H2CNCN}, given the limited number of intense unblended transitions available, we performed its fitting in just one step directly fixing the $\text{FWHM}$ to 10\kms, deriving $N = (9.5 \pm 0.8) \times 10^{11}\, \text{cm}^{-2}$, $T_\text{ex} = 7.6 \pm 0.7$\K and $v_\text{LSR} = 47.5 \pm 0.5$\kms. The derived column densities translate into abundances relative to \ce{H2} of $(3.5 \pm 0.7)\times10^{-10}$ for $Z$-\ce{HNCHCN}, $(7.9 \pm 1.6)\times10^{-11}$ for $E$-\ce{HNCHCN} and $(1.58 \pm 0.34)\times10^{-11}$ for \ce{H2CNCN}.

\begin{figure*}[ht!]
    \centering
    \includegraphics[width=0.98\textwidth]{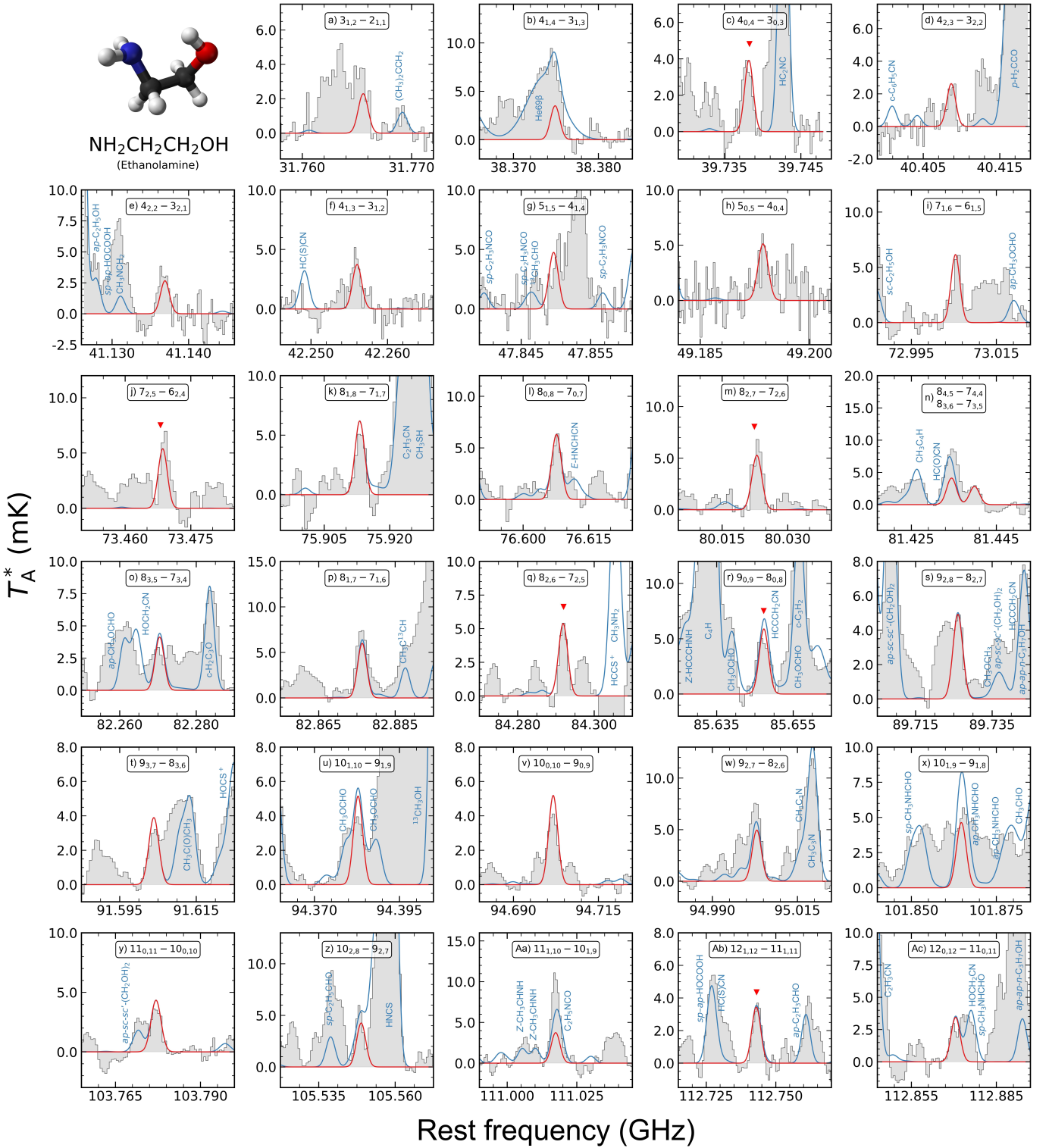}
    \caption{Same as Fig.~\ref{fig:urea_detection}, but for \ce{NH2CH2CH2OH} (ethanolamine). 
    %The LTE analysis for this molecule was done following the two-steps methodology, being those transitions marked with $\blacktriangledown$ symbols employed to constrain the $\text{FWHM}$ in the first step. 
    %The spectroscopic details for all these transitions is given in Table~\ref{tab:NH2CH2CH2OH_rotational_spectroscopy} of Appendix~\ref{appendix:molecular_spectroscopy}. 
    Lines showing a $\blacktriangledown$ symbol were used to constrain the $\text{FWHM}$ in the LTE fit.
    The molecular structure of \ce{NH2CH2CH2OH} is shown in the upper left corner, using the same colours palette as in Fig.~\ref{fig:urea_detection}.
    %\ce{NH2CH2CH2OH} molecular structure is shown in the upper left corner, using the same colours palette as in Fig.~\ref{fig:urea_detection}.
    } 
    \label{fig:NH2CH2CH2OH_detection} 
\end{figure*}

Finally, we report the detection of glycolonitrile (\ce{HOCH2CN}) and $Z$-$1{,}2$-ethenediol ($Z$-\ce{(CHOH)2}) in G+0.633, the former being a fundamental precursor of ribonucleotides, lipids, the amino acids glycine and alanine (e.g., \citealt{Patel2015, Kua&Paradela2020}), and a catalytic agent in the \ce{HCN} oligomerization chain \citep{Schwarzt1982}, while the latter, which is the enol form of glycolaldehyde (\ce{HOCH2CHO}), is believed to be key towards the formation of sugars, one of the essential ingredients of RNA ribonucleotides (e.g., \citealt{Kitadai&Maruyama2018}). 
Thus far, glycolonitrile has only been clearly detected in two star-forming regions: the Solar-type protostar IRAS16293-2422 B \citep{Zeng2019} and Serpens SMM1-a \citep{Ligterink2021}, very recently in the V883 Ori protoplanetary disc \citep{Fadul2025}, and tentatively in G+0.693 \citep{Rivilla2022_nitriles}.
%Thus far, glycolonitrile has only been clearly detected in two star-forming regions: the Solar-type protostar IRAS16293-2422 B \citep{Zeng2019} and Serpens SMM1-a \citep{Ligterink2021}; very recently in the V883 Ori protoplanetary disc \citep{Fadul2025}, and tentatively detected in G+0.693 \citep{Rivilla2022_nitriles}.
%Thus far, glycolonitrile has only been clearly detected towards two star-forming regions: the Solar-type protostar IRAS16293-2422 B \citep{Zeng2019} and Serpens SMM1-a \citep{Ligterink2021}, and tentatively detected towards G+0.693 by \citet{Rivilla2022_nitriles}. 
Meanwhile, $Z$-$1{,}2$-ethenediol has only been reported in G+0.693 \citep{Rivilla2022_ethenediol}. For both species, we significantly expanded the number of detected transitions in G+0.633 in comparison to those reported in G+0.693, shown in Figs.~\ref{fig:HOCH2CN_detection} and \ref{fig:HOCH=CHOH_detection} within Appendix~\ref{appendix:molecules_additional_detection_figures}.
%and we used all of them to perform their LTE fitting. 
Following in both cases the two-step approach, we first derived a $\text{FWHM}$ of 11.2$\pm$1.1\kms for \ce{HOCH2CN} and of 10.3$\pm$1.2\kms for $Z$-\ce{(CHOH)2}. Then, keeping that fixed to these values while leaving the remaining
%other three 
parameters free, we re-run \textsc{AUTOFIT} obtaining $N = (7.4 \pm 0.3) \times 10^{12}\, \text{cm}^{-2}$, $T_\text{ex} = 13.3 \pm 0.9$\K, $v_\text{LSR} = 49.0 \pm 0.3$\kms for \ce{HOCH2CN}, and $N = (9.9 \pm 0.3) \times 10^{12}\, \text{cm}^{-2}$, $T_\text{ex} = 13.0 \pm 0.8$\K, $v_\text{LSR} = 49.0 \pm 0.2$\kms, for $Z$-\ce{(CHOH)2}. We derived abundances relative to \ce{H2} of $(1.23 \pm 0.25)\times10^{-10}$ for \ce{HOCH2CN} and of $(1.65 \pm 0.33)\times10^{-10}$ for $Z$-\ce{(CHOH)2}.
%In the case of \ce{HOCH2CN} we fixed the $v_\text{LSR}$ to 49\kms and the $\text{FWHM}$ to 10\kms for it to converge, deriving $T_\text{ex} = 13 \pm 2$\K and $N = (7.2 \pm 0.7) \times 10^{12}\, \text{cm}^{-2}$. As for $Z$-\ce{(CHOH)2}, we similarly obtained $v_\text{LSR} = 49.8 \pm 0.5$\kms, $T_\text{ex} = 14.0 \pm 4$\K and $N = (9.1 \pm 0.8) \times 10^{12}\, \text{cm}^{-2}$, while keeping fixed the $\text{FWHM}$ to 10\kms. We derived abundances relative to \ce{H2} of $(xx \pm yy)\times10^{-10}$ for \ce{HOCH2CN} and of $(xx \pm yy)\times10^{-10}$ for $Z$-\ce{(CHOH)2}.

\subsection{Lipid and protein precursors: ethanolamine, glycolamide, vinylamine, carbonic acid, ethyl isocyanate and isobutene}

\begin{figure*}[ht!]
    \centering
    \includegraphics[width=0.98\textwidth]{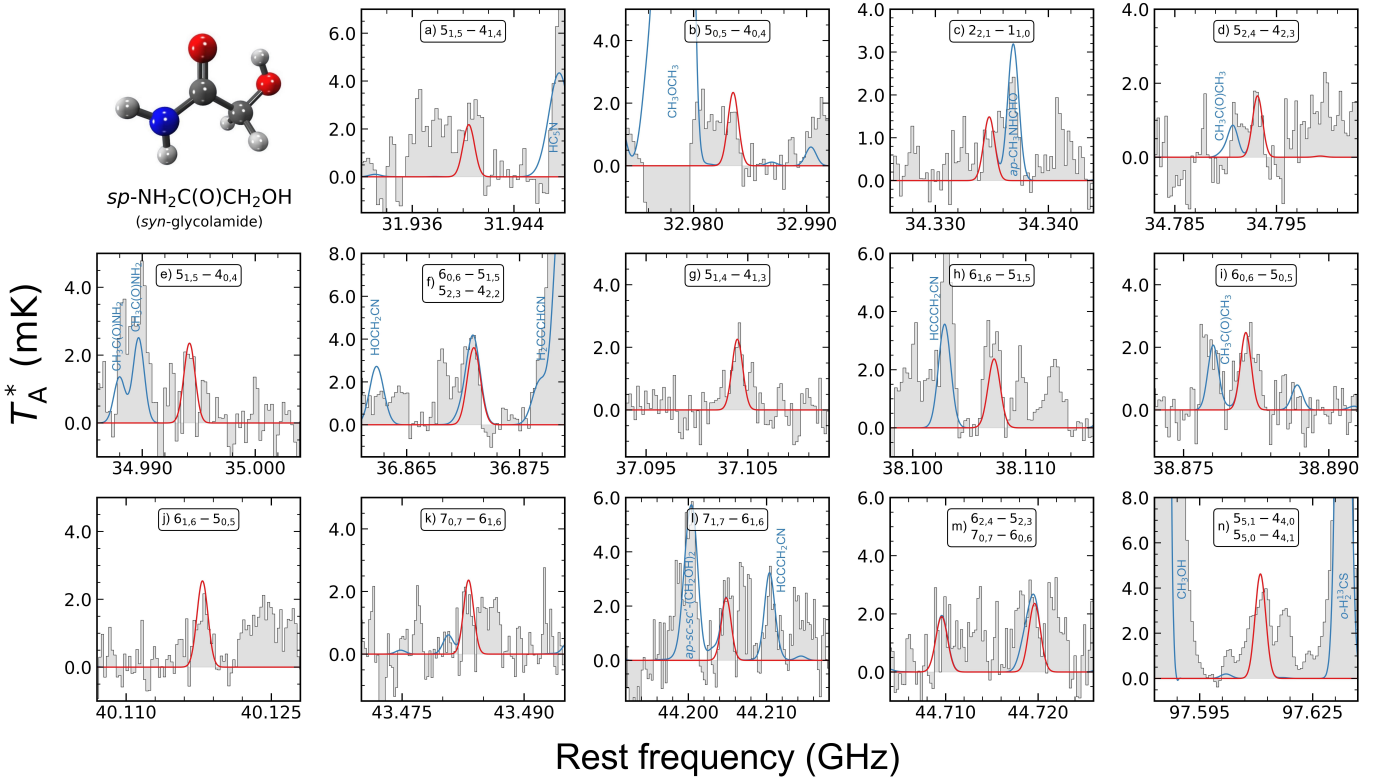}
    \caption{Same as Figs.~\ref{fig:urea_detection}$-$\ref{fig:NH2CH2CH2OH_detection}, but for $sp$-\ce{NH2C(O)CH2OH} ($sp$-glycolamide). 
    %The spectroscopic details for all these transitions is given in Table~\ref{tab:NH2C(O)CH2OH_rotational_spectroscopy} of Appendix~\ref{appendix:molecular_spectroscopy}. 
    The molecular structure of $sp$-\ce{NH2C(O)CH2OH} is shown in the upper left corner, using the same colours palette as in Fig.~\ref{fig:urea_detection}.} 
    \label{fig:NH2C(O)CH2OH_detection} 
\end{figure*}

Ethanolamine (EtA, \ce{NH2CH2CH2OH}) is one of the most significant molecules of prebiotic interest identified so far in the ISM. 
%It is among the most common constituents of the hydrophilic heads of phospholipids, the building blocks of cell membranes \citep{Gordon2003}, 
It constitutes one of the most common hydrophilic head groups of phospholipids, the building blocks of cell membranes (e.g., \citealt{Ali&Szabo2023}), but also a proposed direct precursor of glycine (\ce{NH2CH2COOH}) and alanine (\ce{CH3CH(NH2)COOH}), two of the simplest amino acids \citep{Zhang2017, Wirstrom2007}. Until now, EtA has only been reported towards the G+0.693 molecular cloud by \citet{Rivilla2021_Ethanolamine}, despite multiple searches conducted towards star-forming regions that proved unsuccessful \citep{Widicus2003, Wirstrom2007, Nazari2024}. Here, we present the second robust detection of this species in G+0.633. Fig.~\ref{fig:NH2CH2CH2OH_detection} depicts the 29 brightest and cleanest $R$-branch $a$-type 
%unblended or partially blended 
transitions of EtA present in our survey, which sweep a wide range of energy levels with $3 \leq J_\text{up} \leq 12$, $K_\text{a} \leq 4$ and correspond to all those previously identified in G+0.693 by \citet{Rivilla2021_Ethanolamine}. We performed the LTE analysis for EtA following the two-steps methodology, constraining a $\text{FWHM} = 10.0\pm0.8$\kms by first only fitting the transitions in panels c), j), m), q), r) and Ab) within Fig.~\ref{fig:NH2CH2CH2OH_detection}. The subsequent fit incorporating all transitions in Fig.~\ref{fig:NH2CH2CH2OH_detection}, while keeping the $\text{FWHM}$ fixed to the latter value, 
yields $N = (4.56 \pm 0.17) \times 10^{12}\, \text{cm}^{-2}$, $T_\text{ex} = 13.2 \pm 0.8$\K and $v_\text{LSR} = 47.4 \pm 0.2$\kms.
%as the best estimates. 
We finally derive an abundance for EtA with respect to \ce{H2} of $(7.6 \pm 1.5)\times10^{-11}$.
%We performed the LTE fitting for EtA by using all these transitions and leaving the four parameters free, yielding $N = (4.20 \pm 0.36) \times 10^{12}\, \text{cm}^{-2}$, $T_\text{ex} = 14.0 \pm 2$\K, $v_\text{LSR} = 47.0 \pm 0.4$\kms and $\text{FWHM} = 10.0 \pm 0.8$\kms as the best estimates. We derive an abundance for EtA with respect to \ce{H2} of $(xx \pm yy)\times10^{-10}$.

In the ongoing search for the yet elusive interstellar glycine (\ce{NH2CH2COOH}), the simplest amino acid incorporated into proteins, only one of its structural isomers has been unambiguously identified in the ISM to date. This is the synperiplanar ($sp$, commonly denoted as $syn$) conformer of glycolamide ($sp$-\ce{NH2C(O)CH2OH}), the \ce{C2H5O2N} isomer closest in energy to the two main glycine stereoisomers (lying $\sim$619\K above the most stable conformer I and about $\sim$216\K above conformer II; \citealt{Lattelais2011, Sanz-Novo2019}), and one of the members of this isomeric family with the largest dipole moment, surpassed only by glycine conformer II (4.4\Debye versus 5.5\Debye, respectively; \citealt{Lovas1995}). Thus far, the only unambiguous interstellar detection of $sp$-\ce{NH2C(O)CH2OH} has been reported towards the G+0.693 molecular cloud \citep{Rivilla2023}, while a tentative identification towards the G358.93 MM1 hot core has recently been suggested \citep{Duan2026_glycolamide}.
%Building on the first and, so far, unique unambiguous detection of $sp$-\ce{NH2C(O)CH2OH} in ISM towards G+0.693, as recently reported by \citet{Rivilla2023}, 
Building on these reports, we present here the second unambiguous detection of $sp$-glycolamide in the ISM towards the G+0.633 molecular cloud. Fig.~\ref{fig:NH2C(O)CH2OH_detection} shows all lines of $sp$-\ce{NH2C(O)CH2OH} we have identified in G+0.633, which are almost all those observed by \citet{Rivilla2023} in G+0.693. Since all these lines are rather faint, we performed the LTE analysis for this species in one single step, keeping the $\text{FWHM}$ directly fixed to 10\kms to allow for convergence. We obtained $v_\text{LSR} = 48.7 \pm 0.3$\kms, $T_\text{ex} = 6.0 \pm 0.8$\K and $N = (2.50 \pm 0.16) \times 10^{12}\, \text{cm}^{-2}$, which results in an abundance of $sp$-\ce{NH2C(O)CH2OH} relative to \ce{H2} of $(4.2 \pm 0.9)\times10^{-11}$. 
%We note we have also searched for the less stable $anti$-glycolamide conformer ($\sim$500\K more energetic than its $syn$ counterpart; \citealt{Maris2004}) by using the spectroscopy measured by \citet{Maris2004} and \citet{Sanz-Novo2020_glycolamide}, but did not detect it.

In addition to ethanolamine and $sp$-glycolamide, we also report the second identification in the ISM of vinylamine (also known as ethenamine, \ce{C2H3NH2}) and of the $sp$-$ap$ conformer (commonly known as $cis$-$trans$) of carbonic acid (\ce{HOCOOH}) towards G+0.633, adding to \citet{Zeng2021} and \citet{Sanz-Novo2023} first detections in the G+0.693 cloud, respectively. 
Vinylamine is of prebiotic relevance as a primary amine, a class of molecules carrying the amino (\ce{-NH2}) moiety found in key biomolecules such as amino acids and nucleotides.
%Vinylamine is believed to be  as a key potential precursor in the synthesis of nucleobases and amino acids, since it carries the amino (\ce{-NH2}) moiety. 
Carbonic acid, a central species in numerous biological and geochemical processes on Earth (e.g., oceans pH balance or the global carbon cycle; \citealt{Caldeira&Wickett2003, Wang2016}), 
also stands out as a plausible prebiotic intermediate in carboxylic acid chemistry, a family of compounds widely recognised for their role in prebiotic pathways leading to amino acids and lipids \citep{Kitadai&Maruyama2018, Nogal2023}.
%also stands out as a plausible prebiotic intermediate due to its carboxylic acid character, a family of molecules considered key precursors in the formation of amino acids, proteins and lipids 
%\citep{Ehrenfreund2001,Georgiou&Deamer2014,Smith2015}. 
Figs.~\ref{fig:C2H3NH2_detection} and \ref{fig:cis-trans-HOCOOH_detection} in Appendix~\ref{appendix:molecules_additional_detection_figures} depict all lines of \ce{C2H3NH2} and $sp$-$ap$-\ce{HOCOOH} that are more clearly detected towards G+0.633, respectively. In all cases, we detected in G+0.633 an identical set of transitions as those previously targeted towards G+0.693 in the matching covered spectral ranges (\citealt{Zeng2021} and \citealt{Sanz-Novo2023}, respectively), but expanding on: i) four more energetic ($E_{\text{up}} \sim 80$\K) $J_\text{up} \geq 12$ lines (besides 7 additional lines across the 7\mm and 3\mm windows) for \ce{C2H3NH2} and ii) two lines of $sp$-$ap$-\ce{HOCOOH} that appeared heavily blended in G+0.693 but are only partially blended in G+0.633. We performed the LTE fitting for these two species in just one step, fixing directly the $\text{FWHM}$ to 10\kms given their very limited number of intense unblended lines. The fitting results are $N = (1.2 \pm 0.1) \times 10^{13}\, \text{cm}^{-2}$, $T_\text{ex} = 21.2 \pm 1.5$\K and $v_\text{LSR} = 50.0 \pm 0.3$\kms for \ce{C2H3NH2}; whereas $N = (2.70 \pm 0.23) \times 10^{12}\, \text{cm}^{-2}$, $T_\text{ex} = 8.1 \pm 0.9$\K and $v_\text{LSR} = 49.1 \pm 0.5$\kms for $sp$-$ap$-\ce{HOCOOH}. The resulting abundances relative to \ce{H2} are $(2.0 \pm 0.4)\times10^{-10}$ for \ce{C2H3NH2} and $(4 \pm 1)\times10^{-11}$ for $sp$-$ap$-\ce{HOCOOH}.

\begin{figure*}[ht!]
    \centering
    \includegraphics[width=0.98\textwidth]{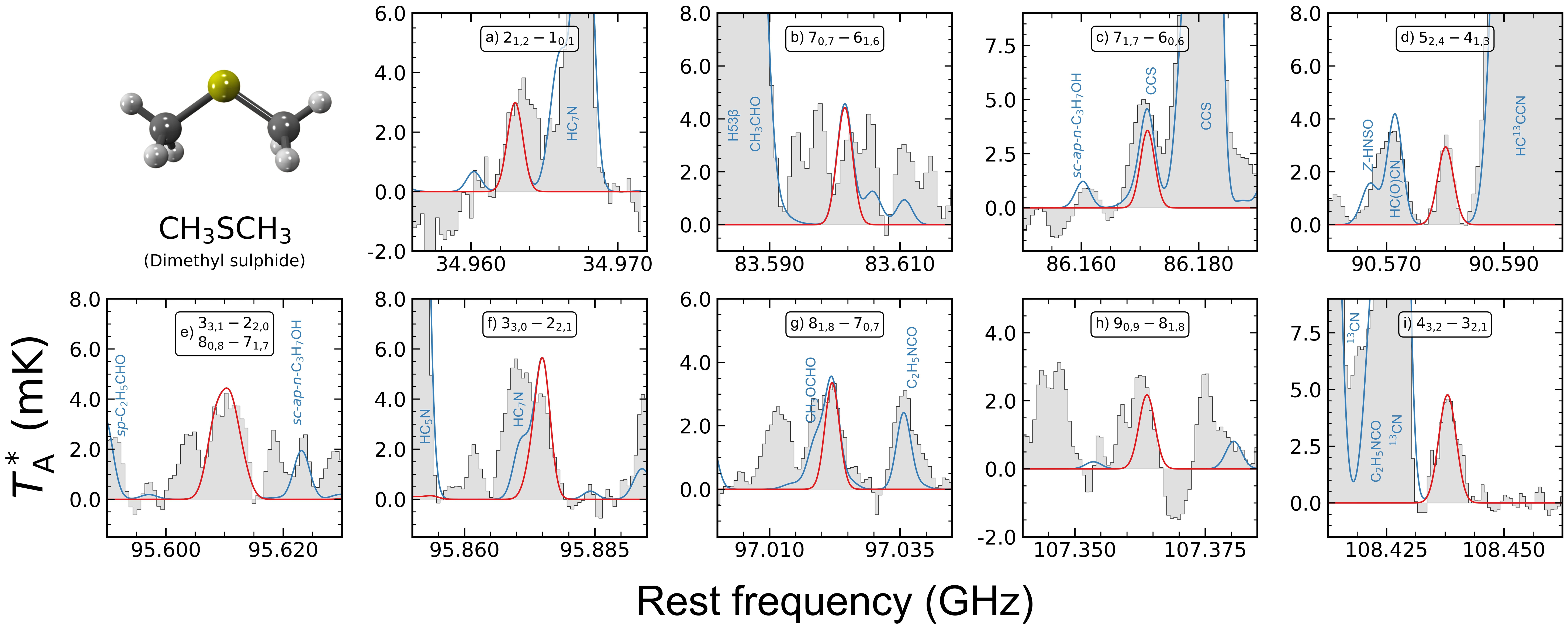}
    \caption{Same as Figs.~\ref{fig:urea_detection}$-$\ref{fig:NH2C(O)CH2OH_detection}, but for \ce{CH3SCH3} (dimethyl sulphide, DMS). 
    %The spectroscopic details for all these transitions is given in Table~\ref{tab:NH2C(O)CH2OH_rotational_spectroscopy} of Appendix~\ref{appendix:molecular_spectroscopy}. 
    The molecular structure of \ce{CH3SCH3} is shown in the upper left corner, using the same colours palette (with sulphur atoms depicted in yellow) as in Fig.~\ref{fig:urea_detection}.}  
    \label{fig:CH3SCH3_detection} 
\end{figure*}

Finally, we also present second reports in the ISM of ethyl isocyanate (\ce{C2H5NCO}) and isobutene (\ce{(CH3)2CCH2}), which are shown in Figs.~\ref{fig:C2H5NCO_detection} and \ref{fig:Isobutene_tentative} of Appendix~\ref{appendix:molecules_additional_detection_figures}, respectively. 
%Ethyl isocyanate is highlighted for its role in peptide polymerisation \citep{Pascal2005}, in addition to the production of nucleotides and nucleosides (e.g., \citealt{Choe2021}). 
%\citealt{Schneider2018}). 
The prebiotic relevance of ethyl isocyanate stems from the proposed role of cyanates in amino acid synthesis, peptide polymerisation, and nucleotides production (e.g., \citealt{Danger2012}; \citealt{Choe2021}).
Meanwhile, \ce{(CH3)2CCH2}, as a branched alkene and one of the few large saturated hydrocarbons identified in space to date, is a potential building block for the long-chain isoprenoid lipids believed to have composed protocell membranes (\citealt{Cleaves2012, Nakatani2012}). Both species have only been identified towards G+0.693 thus far (\citealt{Rodriguez-Almeida2021_C2H5NCO} and \citealt{Fatima2023}, respectively). 
%For both species, the transitions identified in G+0.633 correspond to those observed in G+0.693 by \citet{Rodriguez-Almeida2021_C2H5NCO} and \citet{Fatima2023}, respectively, while including 
Apart from the complete set of transitions previously targeted in G+0.693 by \citet{Rodriguez-Almeida2021_C2H5NCO} and \citet{Fatima2023}, we identified in G+0.633: i) ten more energetic ($E_{\text{up}} \sim 12-55$\K) $K_\text{a} = 0,1$ transitions with $9 \leq J_\text{up} \leq 19$ for \ce{C2H5NCO}, and ii) four additional partially blended lines at 3\mm detected with $\text{S/N} > 5$ in integrated intensity in the case of \ce{(CH3)2CCH2}. 
We note that for \ce{(CH3)2CCH2}, all identified lines are coalesced quartets of transitions resulting from different symmetry substates (A$_1$A$_1$, EE, EA$_1$ and A$_1$E, arising from the internal rotations of the two equivalent \ce{CH3} groups of this species), which cannot be resolved at the linewidths of G+0.633 ($\sim$10\kms). Although the contribution to the observed line profile of some of these coalesced transitions is negligible, we considered all of them to perform the LTE analysis for this species. For both species, we fixed directly the $\text{FWHM}$ to 10\kms.
%we carried out the LTE analysis in just one step, fixing the $\text{FWHM}$ to 10\kms. 
We derived $N = (6.2 \pm 0.9) \times 10^{12}\, \text{cm}^{-2}$, $T_\text{ex} = 24.2 \pm 3.6$\K and $v_\text{LSR} = 48.0 \pm 0.9$\kms for \ce{C2H5NCO}, and $N = (6.7 \pm 0.6) \times 10^{13}\, \text{cm}^{-2}$, $T_\text{ex} = 11.3 \pm 1.8$\K and $v_\text{LSR} = 47.2 \pm 0.4$\kms for \ce{(CH3)2CCH2}. The resulting molecular abundances relative to \ce{H2} are $(1.03 \pm 0.26)\times10^{-10}$ for \ce{C2H5NCO} and $(1.12 \pm 0.24) \times 10^{-9}$ for \ce{(CH3)2CCH2}.

\subsection{\ce{S}-bearing species: second reports in the ISM of DMS (\ce{CH3SCH3}), $Z$-\ce{HNSO} and \ce{HOCS^+}}

Sulphur (\ce{S}) is one of the most abundant elements in the Universe and is of particular relevance for both astrochemistry and astrobiology. In the ISM, \ce{S}-bearing species act as ``chemical clocks'' and tracers of physical conditions, particularly in shock-processed environments (e.g., \citealt{Esplugues2023}), while playing a key role in stellar nucleosynthesis and the chemical evolution of galaxies \citep{Perdigon2021}. Sulphur is also essential for terrestrial biochemistry, being a fundamental constituent of some amino acids and other biomolecules involved in cellular metabolism \citep{Todd2022}, and is believed to be central to several prebiotic chemical networks \citep{Patel2015}.

One of the most significant \ce{S}-bearing species is dimethyl sulphide (DMS, \ce{CH3SCH3}), which has long been proposed as a potential biomarker of aquatic life in habitable exoplanets with \ce{H2}-rich atmospheres (e.g., \citealt{Seager2016, Madhusudhan2021}). However, this interpretation has been challenged by recent astrochemical evidence showing that DMS can also arise through efficient abiotic pathways, following its first unambiguous detection in the ISM towards the G+0.693 molecular cloud \citep{Sanz-Novo2025_DMS} after its identification in the comet $\text{67P/Churyumov-Gerasimenko }$ \citep{Hanni2024}. In this work, we present the second unambiguous report of DMS in the ISM towards G+0.633. Fig.~\ref{fig:CH3SCH3_detection} compiles the clearest DMS transitions within G+0.633 survey, which constitute a reduced subset of those targeted in G+0.693 by \citet{Sanz-Novo2025_DMS}. Similarly to \ce{(CH3)2CCH2}, all \ce{CH3SCH3} detected lines in G+0.633 correspond to coalesced quartets of transitions related to different symmetry substates (AA, EE, EA and AE), which cannot be resolved at the $\sim$10\kms linewidths of G+0.633. To perform the fit for this species, we also considered all of the coalesced transitions constituting each line regardless of their contribution to the observed profile, and kept fixed $\text{FWHM} = 10$\kms. We obtained $v_\text{LSR} = 49.9 \pm 0.4$\kms and $T_\text{ex} = 7.4 \pm 0.6$\K, $N = (1.03 \pm 0.06) \times 10^{13}, \text{cm}^{-2}$ and a $N/N_{\ce{H2}}$ abundance of $(1.7 \pm 0.4) \times 10^{-10}$. This second report of DMS in G+0.633 further challenges its reliability as an unambiguous biosignature again.
%, underscoring the need to carefully assess the potential abiotic formation of DMS in exoplanetary studies.
%This second report of DMS in G+0.633 further challenges its interpretation as a plausible biosignature, underscoring the need to carefully assess the potential abiotic formation of DMS in exoplanetary studies.
%Given the scarce number of completely unblended \ce{CH3SCH3} transitions identified in G+0.633, and the relatively modest S/N at which they are detected,
%along with the large uncertainties related to the fitting, we opted to classify the presence of \ce{CH3SCH3} in G+0.633 as tentative, yet highly probable.

In addition to DMS, we also report the second interstellar detection of the $Z$ stereoisomer (commonly denoted as $cis$) of thionylimide (\ce{HNSO}) towards G+0.633, the most stable isomer among the simplest [H,N,S,O] representatives \citep{Bharatam2006}. This follows its very recent first identification towards G+0.693 by \citet{Sanz-Novo2024_HNSO}. \ce{NSO} compounds (i.e., molecules that simultaneously contain \ce{N}, \ce{S} and \ce{O}) are species of great interest on Earth, both in geosciences, given their role as tracers of biotic and paleoenvironmental signatures in sedimentary rocks (e.g., \citealt{Ji2021}),
%; \citealt{Yue2023}), 
but also from the biological perspective, as they are believed to be key in the transmission of signals within and between cells and tissues (e.g., \citealt{Miljkovic2013}). These species are of particular astrochemical interest because they trace the chemical interplay between \ce{N}-, \ce{S}-, and \ce{O}-bearing reservoirs \citep{delValle2026}.
%However, their search is space has barely taken place.
Fig.~\ref{fig:cis-HNSO_detection} of Appendix~\ref{appendix:molecules_additional_detection_figures} shows all 
%unblended or slightly blended 
lines of $Z$-\ce{HNSO} that we have clearly identified in G+0.633, which correspond to various $K_\text{a} = 0,1,2$ transitions arising from $J_\text{up} = 2, 4, 5$ energy levels, all of which have also been targeted in G+0.693. 
Following the two-step approach, we first estimate a $\text{FWHM}$ of 11.2$\pm$0.9\kms by only fitting the most intense unblended transitions depicted in panels f), i) and k) within Fig.~\ref{fig:cis-HNSO_detection}. Keeping the $\text{FWHM}$ fixed to this value, we then applied \textsc{AUTOFIT} again incorporating all transitions shown in Fig.~\ref{fig:cis-HNSO_detection}, yielding $v_\text{LSR} = 49.2 \pm 0.3$\kms, $T_\text{ex} = 12.1 \pm 1.3$\K, $N = (2.1 \pm 0.1)\times10^{13}\, \text{cm}^{-2}$ and $N/N_{\ce{H2}} = (3.5 \pm 0.7)\times10^{-10}$.
%We performed the LTE fitting for this molecule by using all these transitions and leaving all four physical parameters free when running AUTOFIT, yielding $N = (1.98 \pm 0.16)\times10^{13}\, \text{cm}^{-2}$, $T_\text{ex} = 9.2 \pm 1.4$\K, $v_\text{LSR} = 49.2 \pm 0.4$\kms and $\text{FWHM} = 11.0 \pm 0.9$\kms. The derived $Z$-\ce{HNSO} abundance relative to \ce{H2} is $(3.5 \pm 0.7)\times10^{-10}$.

Finally, following the first report of \ce{HOCS+} towards G+0.693 by \citet{Sanz-Novo2024_HOCS+}, we present its second interstellar detection in G+0.633. This \ce{S}-bearing cation constitutes a valuable probe of sulphur chemistry in the ISM, particularly of ion-molecule reactions involved in the formation of more complex \ce{S}-bearing species with potential prebiotic relevance (e.g., \citealt{Parker2011, Todd2022}). We detected the six least energetic $K_\text{a} = 0$ lines among those observed in G+0.693 by \citet{Sanz-Novo2024_HOCS+}, together with three intense $K_\text{a} = 1$ lines later reported by \citet{Lattanzi2024}, which are shown in Fig.~\ref{fig:HOCS+_tentative} within Appendix~\ref{appendix:molecules_additional_detection_figures}. As in G+0.693, the $K_\text{a} = 0$ lines are partially blended with the emission from 
%the $a$-type spectrum of the \ce{^{34}S} isotopologue of \ce{HNCS} (i.e., \ce{HNC^{34}S}), 
\ce{HNC^{34}S} (see \citealt{Sanz-Novo2024_HOCS+} for further details); 
%but whose contribution in G+0.633 can be more readily disentangled from \ce{HOCS+} profiles owing to its comparatively $\sim$2 times narrower linewidths (see Sect.~\ref{sec:G0633_vs_G0693}). 
however, the comparatively $\sim$2 times narrower linewidths observed in G+0.633 allow both species to be more readily disentangled.
We therefore performed a simultaneous fit of \ce{HOCS+} and \ce{HNC^{34}S}, fixing $v_\text{LSR} = 48.5$\kms and $\text{FWHM} = 10$\kms, while adopting $T_\text{ex} = 28$\K for \ce{HOCS+} and $T_\text{ex} = 20.4$\K for \ce{HNC^{34}S}, as derived in G+0.693.
%Therefore, and given that some $K_\text{a} = 1$ transitions not blended with \ce{HNC^{34}S} are also identified in G+0.633, we performed directly the fitting of \ce{HOCS+} and \ce{HNC^{34}S} simultaneously by i) fixing $v_\text{LSR} = 48.5$\kms and $\text{FWHM} = 10$\kms for both molecules;
%, attending to the best estimates obtained for most of the species identified in G+0.633; 
%and ii) fixing $T_\text{ex} = 28$\K for \ce{HOCS+} and $T_\text{ex} = 20.4$\K for \ce{HNC^{34}S}, as found towards G+0.693 by \citet{Sanz-Novo2024_HOCS+} and based on G+0.693 and G+0.633 resemblance in excitation conditions (Sect.~\ref{sec:G0633_vs_G0693}). 
We obtained $N = (2.08 \pm 0.15) \times 10^{12}\, \text{cm}^{-2}$ for \ce{HOCS+} and $N = (5.9 \pm 1.2) \times 10^{11}\, \text{cm}^{-2}$ for \ce{HNC^{34}S}, which translate into molecular abundances relative to \ce{H2} of $(3.5 \pm 0.7) \times 10^{-11}$ and $(1.0 \pm 0.3) \times 10^{-11}$, respectively. This becomes the second detection of \ce{HNC^{34}S} in the ISM. The resulting \ce{HOCS+}/\ce{HNC^{34}S} ratio (3.5$\pm$0.6) is consistent with that reported for G+0.693 ($\sim$3; \citealt{Sanz-Novo2024_HOCS+}), validating the adopted methodology.

%__________________________________________________________________

\begin{figure*}[ht!]
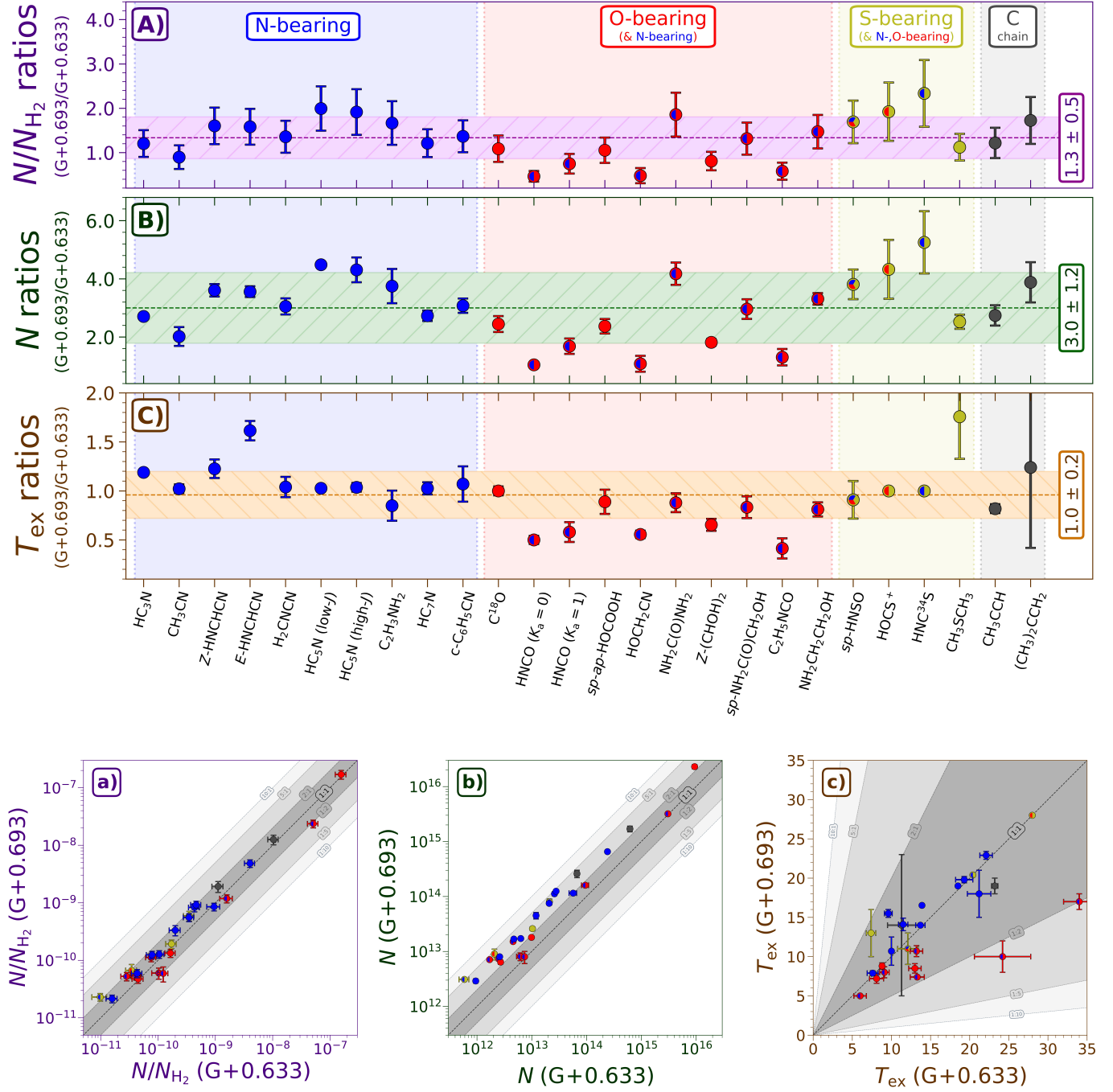

    \centering
    \begin{subfigure}[b]{\textwidth}
        \centering
        \includegraphics[width=0.98\textwidth]{G0693_vs_G0633_Abundances_N_Tex_comparison_ratios.jpeg}
    \end{subfigure}
    %\noindent\hrulefill
    %\\[-0.8ex]
    %\noindent\hdashrule{\textwidth}{0.5pt}{4pt 2pt}
    %\\[0.4ex]
    \\[2.5ex]
    \begin{subfigure}[b]{\textwidth}
        \centering
        \includegraphics[width=0.98\textwidth]{G0693_vs_G0633_Abundances_N_Tex_comparison_absolute_values.jpeg}
    \end{subfigure}
    \caption{Comparison of excitation temperatures ($T_{\text{ex}}$), column densities ($N$) and molecular abundances ($N/N_{\ce{H2}}$) observed in G+0.693 versus G+0.633, for all \ce{N}-, \ce{O}-, \ce{S}-bearing and carbon chain species reported in G+0.633 so far (\citealt{Rivilla2026_benzo}, Paper I and this work). \ce{N}-bearing molecules are shown in blue, \ce{O}-bearing in red (with added blue if containing \ce{N}), \ce{S}-bearing in olive (with added blue and/or red to indicate the presence of \ce{N} and/or \ce{O}, respectively), and \ce{C} chains in dark grey. 
    %The figure presents the same three quantities in two complementary ways. 
    Top (A-C): relative $N/N_{\ce{H2}}$, $N$ and $T_\text{ex}$ ratios of G+0.693 over G+0.633, with the violet, green and orange dashed lines representing the respective average values. The grated regions in similar colours indicate the corresponding 1$\upsigma$ uncertainty intervals. The values of these mean ratios and their 1$\upsigma$ uncertainties are provided in the boxes on the right. Bottom (a-c): correlation diagrams between the $N/N_{\ce{H2}}$, $N$ and $T_\text{ex}$ derived in G+0.693 with respect to those retrieved in G+0.633 for the same set of species. Contours defining the linear relations $y = nx$, with $n = 1; 1/2, 2; 1/5, 5; 1/10, 10$ are shadowed in a fading gradient of gray tones, with the reciprocal relations sharing the same colour.} \label{fig:G0693_G0633_Abundances_N_Tex_comparison_all} 
\end{figure*}

\section{Discussion}
\label{sec:discussion}

%\subsection{G+0.633 v.s. G+0.693: a duel of astrochemical titans}
%\subsection{Comparison between the G+0.693 and G+0.633 clouds}
\label{sec:G0633_vs_G0693}

\begin{figure*}[ht!]
    \centering
    \begin{subfigure}[b]{0.345\textwidth}
        \centering
        \includegraphics[width=0.98\textwidth]{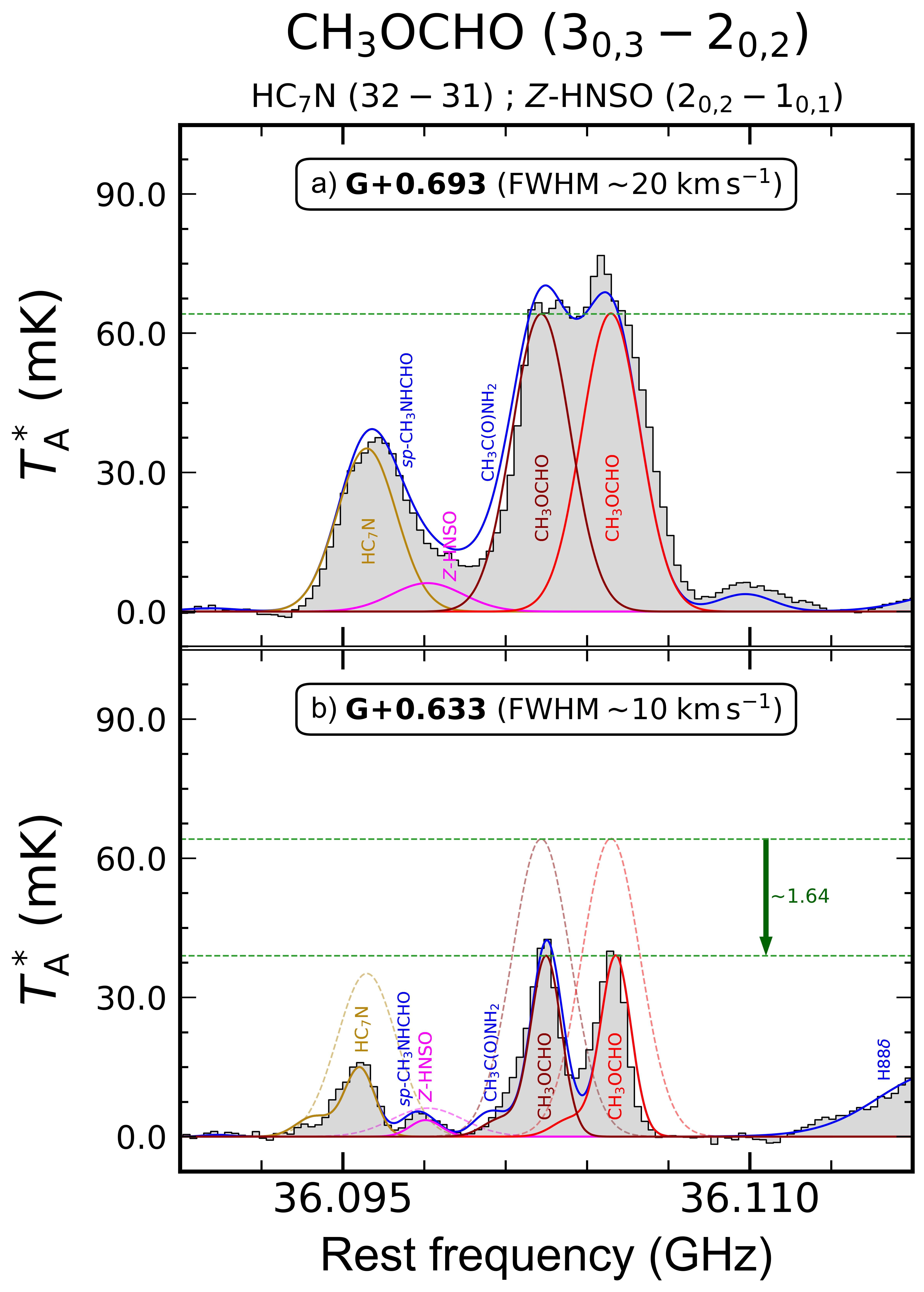}
    \end{subfigure}
    \begin{subfigure}[b]{0.32\textwidth}
        \centering
        \includegraphics[width=0.98\textwidth]{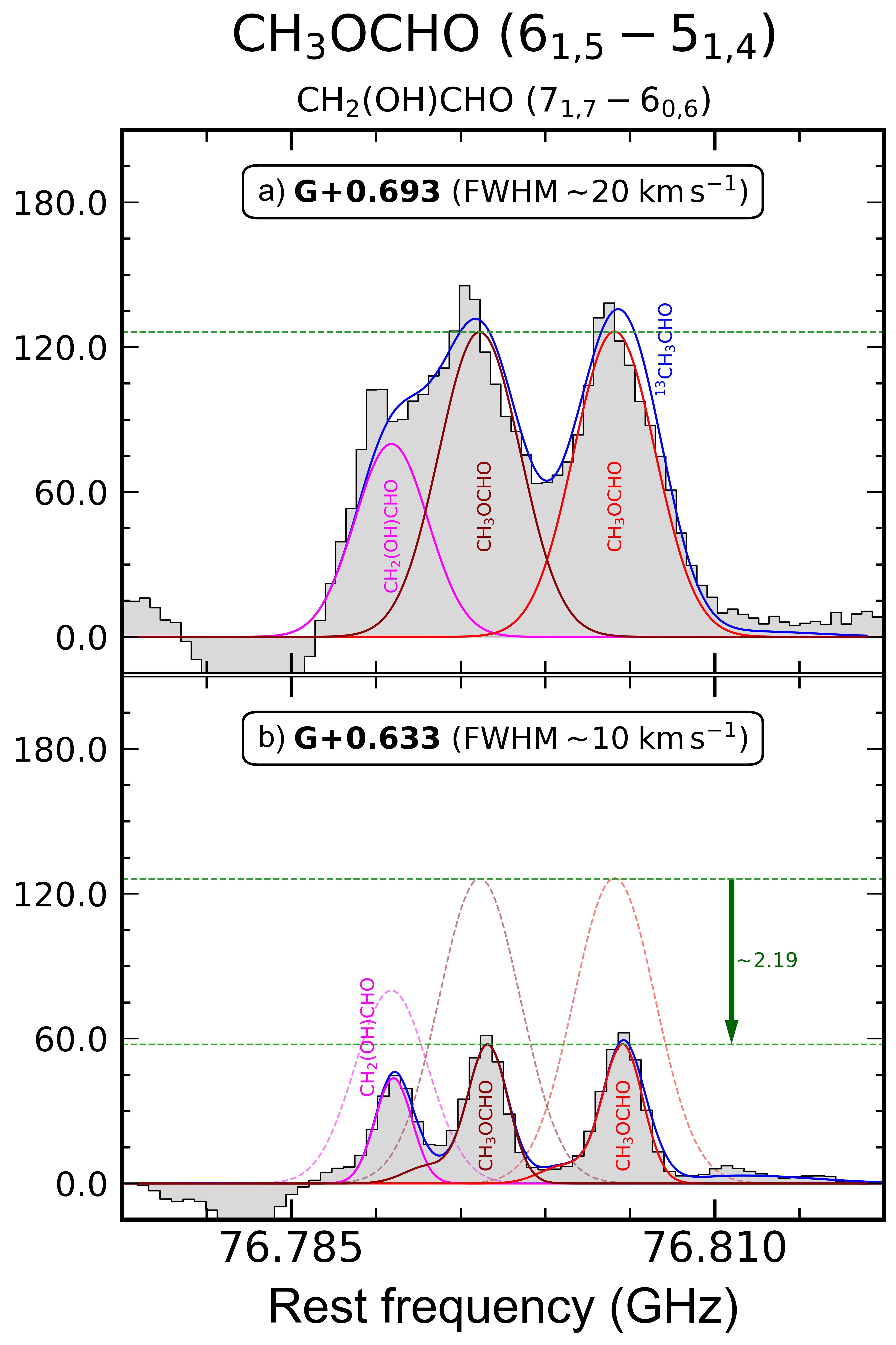}
    \end{subfigure}
    \begin{subfigure}[b]{0.32\textwidth}
        \includegraphics[width=0.98\textwidth]{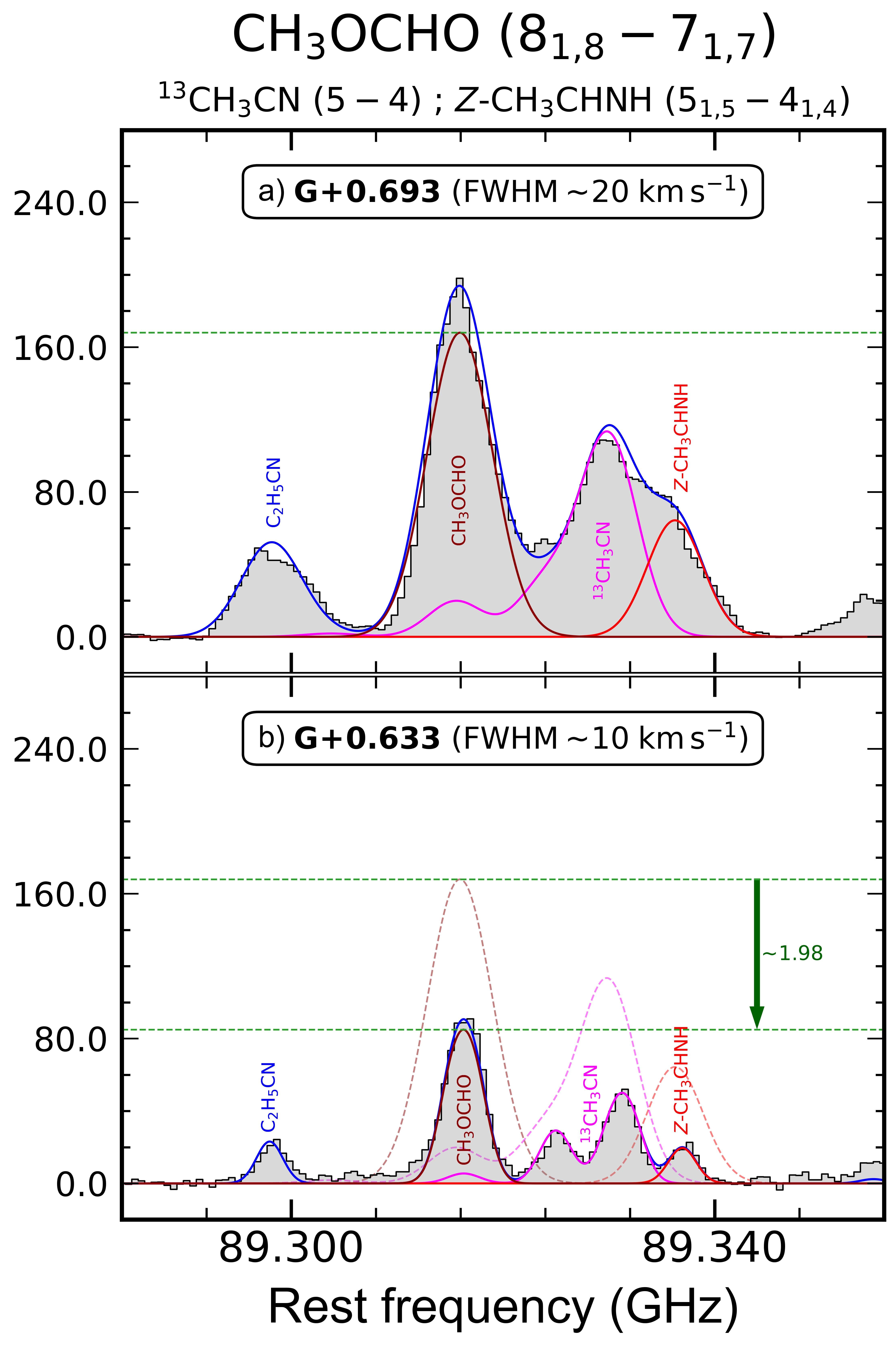}
    \end{subfigure}
    \caption{Comparison of the widths and relative intensities (green lines, arrows and labels) of distinct sets of spectral lines in G+0.693 (upper panels) versus G+0.633 (lower ones). The grey-shaded black histograms indicate the observed spectrum, while the red, brown, magenta and amber solid lines represent the synthetic LTE spectrum of the molecules whose emission profile is unblended in G+0.633, but not in G+0.693 (with their respective label). The title of each subplot indicates the respective rotational transition(s). Their combined profiles, along with the emission from the rest of molecules so far characterised in these clouds, are shown in blue.}
    \label{fig:G0693_G0633_FWHM_line_intensities_comparison}
\end{figure*}

On the basis of its remarkable chemical richness, the G+0.633 molecular cloud directly ranks as one of the most prolific astrochemical laboratories, with a potential comparable to
%, if not exceeding, 
that of the prototypical G+0.693 cloud. Given their spatial proximity, analogous environmental conditions (Paper I) and nearly identical molecular feedstocks (as seen from previous sections), these two clouds offer an excellent framework for a detailed comparative study. In this section, we compare their observed molecular emission properties, %as traced by their observed molecular emission, 
%in a pulse of strength to 
in order to advance the search for new complex species in the ISM. We note that this comparison has been done by only considering the main velocity component observed in both G+0.633 (Paper I) and G+0.693 \citep{Colzi2024}.

One of the first notable features are the similar gas temperature and excitation conditions that these two clouds share. This is clearly illustrated in Fig.~\ref{fig:G0693_G0633_Abundances_N_Tex_comparison_all}, which provides a comparison of the $T_\text{ex}$ derived for 
%40 selected
all \ce{N}-, \ce{O}-, \ce{S}-bearing and carbon chain species that are detected in both clouds and so far reported in G+0.633 (\citealt{Rivilla2026_benzo}; Paper I), including all those presented in this work. For most of the molecules, the derived $T_\text{ex}$ in both G+0.693 and G+0.633 are typically found to be $\sim$5$-$20\K, 
%(see, e.g., \citealt{Rivilla2020a, Colzi2022, SanAndres2023} for G+0.693; Table~\ref{tab:detected_molecules_fitting_parameters} for G+0.633)
resulting into similar $T_\text{ex}$ ratios close to unity among the same species (Fig.~\ref{fig:G0693_G0633_Abundances_N_Tex_comparison_all}). Moreover, this trend holds across different molecular families and their isotopologues, though more dispersion is observed for some \ce{O}-bearing species. 
%Nonetheless, the observed dispersion (not only associated with \ce{O}-bearing molecules) could be attributable to non-LTE effects not considered when performing their fittings, or even to a reduced number of transitions available for some species (either within G+0.633 or G+0.693 surveys) when published (e.g., \ce{H2CNCN} or \ce{CH3SCH3} in G+0.633; \ce{C2H5NCO} or \ce{HOCH2CN} in G+0.693), and might therefore be mitigated when reevaluating the analysis of these molecules using more extensive and sensitive datasets. 
The observed scatter (not only among \ce{O}-bearing molecules) may primarily arise from the limited number of transitions available for some species in either the G+0.633 or G+0.693 surveys at the time of their analysis (e.g., \ce{H2CNCN} or \ce{CH3SCH3} in G+0.633; \ce{C2H5NCO} or \ce{HOCH2CN} in G+0.693). Non-LTE effects not considered in the fittings may also contribute to the dispersion. Both effects could be alleviated through a reanalysis of these species using more extensive and sensitive datasets.
Notably, the very low $T_\text{ex}$ contrasts sharply with the much higher $T_\text{kin}$ measured in both sources, of $\sim$70$-$140\K for G+0.693 (\citealt{Zeng2018, Colzi2024}) and slightly lower of $\sim$55$-$90\K for G+0.633 (Paper I), indicating subthermal excitation in both cases. This favours spectral line identification as molecular emission is restricted to the low-energy transitions, making molecular lines less susceptible to contamination from overlapping features. Besides the low $T_\text{ex}$, both clouds also exhibit very similar and rather low dust temperatures ($T_\text{dust}$) of $\sim$20\K \citep{Etxaluze2013, Battersby2025}, which indicates that no thermal coupling is present between the gas and the dust (see also Paper I).
%This establishes both G+0.693 and G+0.633 as ideal targets in the search for molecules of greater complexity.

Despite their similar temperature regimes, column densities in G+0.633 are $\sim$3$-$4 times lower than those reported in G+0.693 for most species, as also depicted in Fig.~\ref{fig:G0693_G0633_Abundances_N_Tex_comparison_all}. Nonetheless, molecular abundance ratios between species and both spatial and structural isomers appear to be conserved between both clouds, thus indicating similar chemical frameworks. Taking the species analysed in this work as a proxy, this is evidenced through the $Z/E$ ratio of $C$-cyanomethanimine (\ce{HCNHCN}) stereoisomers and the ratio of the $C$- and $N$-cyanomethanimine structural isomers, being 4.4$\pm$0.3 and 26.9$\pm$2.6 in G+0.633 versus 4.5$\pm$0.2 and 31.8$\pm$1.8 in G+0.693 \citep{SanAndres2024}, respectively. 
%Moreover, we derive a $sc$-$ap$/$ap$-$ap$-$n$-\ce{C3H7OH} ratio of 1.1$\pm$0.1, versus 1.6$\pm$0.2 in G+0.693 \citep{Jimenez-Serra2022}.
%and a $s/a$-\ce{H2CCHOH} ratio of 53$\pm$10 (versus 1.64 and 8.5, respectively; \citealt{Jimenez-Serra2022}). 
Besides, the \ce{H2} column density derived in G+0.693 is only about twice that of G+0.633 (1.35$\times10^{23}\, \text{cm}^{-2}$ versus 0.6$\times10^{23}\, \text{cm}^{-2}$; \citealt{Martin2008}; Paper I, respectively), ultimately resulting in comparable (although generally lower for G+0.633 by a factor of $\sim$1.5) fractional abundances for all species in these two sources (see Fig.~\ref{fig:G0693_G0633_Abundances_N_Tex_comparison_all}). Consequently, the G+0.693 and G+0.633 clouds can be established as a powerful benchmarking pair for astrochemical studies, suitable not only for confirming the detection of new interstellar species, but also for the validation and refinement of astrochemical models. More broadly, the strong correlations in molecular abundances observed between G+0.633 and G+0.693, consistent with recent comparative studies across distinct Galactic shock-dominated environments \citep{Lopez-Gallifa2025}, reinforce the emerging picture of a remarkably homogeneous chemistry in shock-dominated regions, within which the extraordinary molecular richness of these two clouds emerges as a natural outcome.

Beyond the similarities between the two clouds, the G+0.633 cloud exhibits a particularly compelling observational advantage over G+0.693, providing a major opportunity
%which could notably 
to advance the identification of new interstellar species. This advantage comes from its significantly narrower linewidths of $\sim$10\kms, in stark contrast to the much broader $\sim$18$-$25\kms lines observed in G+0.693 (e.g., \citealt{Rivilla2022_nitriles}; \citealt{Zeng2018}), as shown in Fig.~\ref{fig:G0693_G0633_FWHM_line_intensities_comparison}. 
%This substantial reduction in linewidth effectively mitigates line blending, thereby enabling the identification of multiple unblended transitions in the spectra, a key aspect towards the secure detection of large complex species which are expected to exhibit very low abundances and hence much weaker emission lines. 
This substantial reduction in linewidth effectively mitigates line blending, thereby enabling the identification of multiple unblended transitions in the spectra, a key requirement for the secure detection of increasingly complex molecules. Besides their expected low abundances, such species exhibit progressively larger rotational partition functions, which spread the molecular population over a greater number of energy levels and consequently result in much weaker emission lines.
This advantage is clearly evidenced in Fig.~\ref{fig:G0693_G0633_FWHM_line_intensities_comparison}, which highlights how many of the particularly line-crowded spectral features that appear blended in G+0.693 are resolved into isolated lines in G+0.633, resulting in much cleaner spectra and more robust detections despite the fainter profiles. Consequently, the line-confusion limit is crucially reduced, improving the effective detection threshold for molecules with increasing complexity. 
%This crucially lowers line-confusion limits and enhances the effective detection threshold, thus facilitating the identification of molecules with increasing complexity. 
%Furthermore, narrower lines allow for a more precise determination of molecular emission physical parameters of, leading to more stringent estimates for $T_\text{ex}$, column densities and abundance ratios, which are essential for constraining the chemistry of the observed species under interstellar conditions. 
The reduced line blending afforded by narrower linewidths enables more reliable measurements of $T_\text{ex}$, column densities, and abundance ratios, providing stronger constraints on the chemistry of the observed species under interstellar conditions.
Moreover, the half-narrower linewidths also compensate the effect of the $\sim$3$-$4 times lower column densities observed in G+0.633, limiting the reduction of spectral lines peak intensities to only a factor of $\sim$2 (see Fig.~\ref{fig:G0693_G0633_FWHM_line_intensities_comparison}). Although deeper integrations will then be required for G+0.633 to reach the same detection significance (i.e., S/N) achieved in G+0.693, the sensitivity improvements needed are readily achievable with current instruments.
%and modest extra observational effort. 
As observational capabilities improve, the combination of high chemical richness and reduced line confusion is expected to make G+0.633 a premier target for ultra-deep astrochemical surveys, enabling the discovery of new and increasingly complex molecules.
%thus making G+0.633 surveys worthwhile.

Overall, the G+0.633 cloud emerges as a powerful astrochemical laboratory for future studies of molecular complexity in the ISM. Further observations, particularly at the still unexplored 2 and 1\mm bands, will be essential to fully exploit its potential, strengthening the current molecular inventory and enabling the discovery of new species. Among them are metal-bearing molecules such as \ce{MgS}, \ce{NaS} and \ce{CaO}, already reported towards G+0.693 \citep{Rey-Montejo2024}, as well as deuterated species including \ce{DCN}, \ce{DNC}, \ce{DCO+} and \ce{N2D+}, whose identification remains inconclusive in the present survey. Their analysis will provide valuable insights into deuterium fractionation in G+0.633, offering a key diagnostic of the early stages of star formation in this remarkable GC source (e.g., \citealt{Colzi2022}). Future observations at high angular resolution will also be key to resolving the spatial distribution of these and chemically related species, providing new constraints on their chemistry.

%Overall, the G+0.633 cloud emerges as a powerful astrochemical laboratory for future investigations of molecular complexity in the ISM. To fully exploit its potential, further observations are required, particularly at the still unexplored 2 and 1\mm bands. These additional surveys will not only strengthen the current molecular inventory, but also provide new opportunities to discover new species. Among them are metal-bearing molecules such as \ce{MgS}, \ce{NaS} and \ce{CaO}, already reported towards G+0.693 \citep{Rey-Montejo2024}, as well as deuterated species including \ce{DCN}, \ce{DNC}, \ce{DCO+} and \ce{N2D+}, whose identification remains inconclusive in the present survey. Their analysis will provide valuable insights into deuterium fractionation in G+0.633, offering a key diagnostic of the early stages of star formation in this unique GC source (e.g., \citealt{Colzi2022}).

%__________________________________________________________________

\section{Summary and conclusions}
\label{sec:summary_and_conclusions}

We have presented the first detailed chemical characterisation of the newly identified GC G+0.633-0.0604 molecular cloud, a highly chemically rich source located in the southernmost part of the Sgr B2 complex. This work is based on an unprecedented, ultra-deep, single-dish molecular line survey covering $\sim$100\GHz in aggregated bandwidth across the 7, 3, and 1.3\mm bands, performed with the Yebes 40m, IRAM 30m, and APEX radio telescopes.
We report the detection of 15 molecules of high prebiotic and astrochemical relevance, embedded within a census of more than 120 species (about half of which are COMs) and nearly half as many isotopologues. These include urea (\ce{NH2C(O)NH2}), ethanolamine (\ce{NH2CH2CH2OH}), glycolamide ($sp$-\ce{NH2C(O)CH2OH}), dimethyl sulphide (\ce{CH3SCH3}), cyanomethanimines ($Z,E$-\ce{HCNHCN}, \ce{H2CNCN}), glycolonitrile (\ce{HOCH2CN}), $Z$-1,2-ethenediol ($Z$-\ce{(CHOH)2}), vinylamine (\ce{C2H3NH2}), carbonic acid ($sp$-$ap$-\ce{HOCOOH}), ethyl isocyanate (\ce{C2H5NCO}), isobutene (\ce{(CH3)2CCH2}), thionylimide ($Z$-\ce{HNSO}) and \ce{HOCS^+}. Many of these species are reported here for the first time outside the prototypical chemically-rich GC G+0.693-0.027 cloud, establishing G+0.633 as the first confirmed ``astrochemical twin'' of G+0.693.
Moreover, our results demonstrate that the exceptional molecular complexity observed in G+0.693 is not unique, but rather a more natural outcome of shock-processed chemistry in the GC, suggesting that its apparent rarity may largely reflect observational biases due to the limited number of deeply surveyed sources.
%but likely more widespread across the GC. 
G+0.633 shows striking similarities with G+0.693 in excitation conditions, molecular abundances, and overall chemical richness, although its column densities are typically a factor of $\sim$3$-$4 lower. Importantly, G+0.633 benefits from significantly narrower linewidths ($\sim$10\kms versus $\sim$20\kms in G+0.693), which strongly reduces line confusion and makes it an exceptional laboratory for molecular line identification. This places G+0.633 as a uniquely powerful target for future deep spectral surveys and high-angular-resolution observations, optimally suited to search for new molecules of increasing complexity while providing stringent constraints on the chemistry of shock-dominated environments in the GC.

\begin{acknowledgements}

%We thank the anonymous reviewer for the constructive report that improved this manuscript. 
We thank the anonymous reviewer for the constructive report that improved this manuscript. We are very grateful to the Yebes 40 m, IRAM 30 m, and APEX telescope staff for their support during the observing runs. The 40m radio telescope at Yebes Observatory is operated by the Spanish Geographic Institute (IGN; Ministerio de Transportes, Movilidad y Agenda Urbana). IRAM is supported by INSU/CNRS (France), MPG (Germany) and IGN (Spain). APEX is a collaboration between the Max-Planck-Institute für Radioastronomie, the European Southern Observatory, and the Onsala Space Observatory. D.S.A. expresses his gratitude for the financial support provided by the Comunidad de Madrid through the Grant PIPF-2022/TEC-25475. D.S.A. and V.M.R. acknowledge the funds provided by the Consejo Superior de Investigaciones Cient{\'i}ficas (CSIC) and the Centro de Astrobiolog{\'i}a (CAB) through the project 20225AT015 (Proyectos intramurales especiales del CSIC). D.S.A., V.M.R., and M.S.-N. are grateful for the financial support provided from grant CNS2023-144464 funded by MICIU/AEI/10.13039/501100011033 and by ``European Union NextGenerationEU/PRTR''. D.S.A., V.M.R., L.C., M.S.-N., and I.J.-S. acknowledge support from the grant PID2022-136814NB-I00 funded by the Spanish Ministry of Science, Innovation and Universities/State Agency of Research MICIU/AEI/10.13039/501100011033 and by ERDF, UE. V.M.R. also acknowledges support from grant RYC2020-029387-I funded by MICIU/AEI/10.13039/501100011033 and by ``ESF, Investing in your future''. M.S.-N. acknowledges an Alexander von Humboldt postdoctoral fellowship from the Alexander von Humboldt Foundation, and funding from the Marie Sklodowska Curie fellowship SOUL (project number 101272259) funded by the European Union.
L.C., I.J.-S., and S.Z. acknowledge support from the CSIC Bilateral project SOULMATE (project BIJSP25017).
I.J.-S. also acknowledges support by ERC grant OPENS, GA No. 101125858, funded by the European Union. Views and opinions expressed are however those of the author(s) only and do not necessarily reflect those of the European Union, the European Commission, or the European Research Council Executive Agency. Neither the European Union nor the granting authority can be held responsible for them. 
S.Z. acknowledges the support by RIKEN Special Postdoctoral Researchers Program. The project that gave rise to these results received the support of a fellowship from the ``la Caixa'' Foundation (ID 100010434). The fellowship code is LCF/BQ/PR25/12110012.

\end{acknowledgements}


\bibliography{biblio}{}
\bibliographystyle{aa}

\onecolumn
\begin{appendix}

%\section{Details on the achieved sensitivity (rms) within G+0.633 observational molecular line survey}
\section{Overview of G+0.633 molecular line survey: spectral coverage and sensitivity}
\label{appendix:rms}

\begin{figure*}[ht!]
    \centering
    \includegraphics[width=0.98\textwidth]{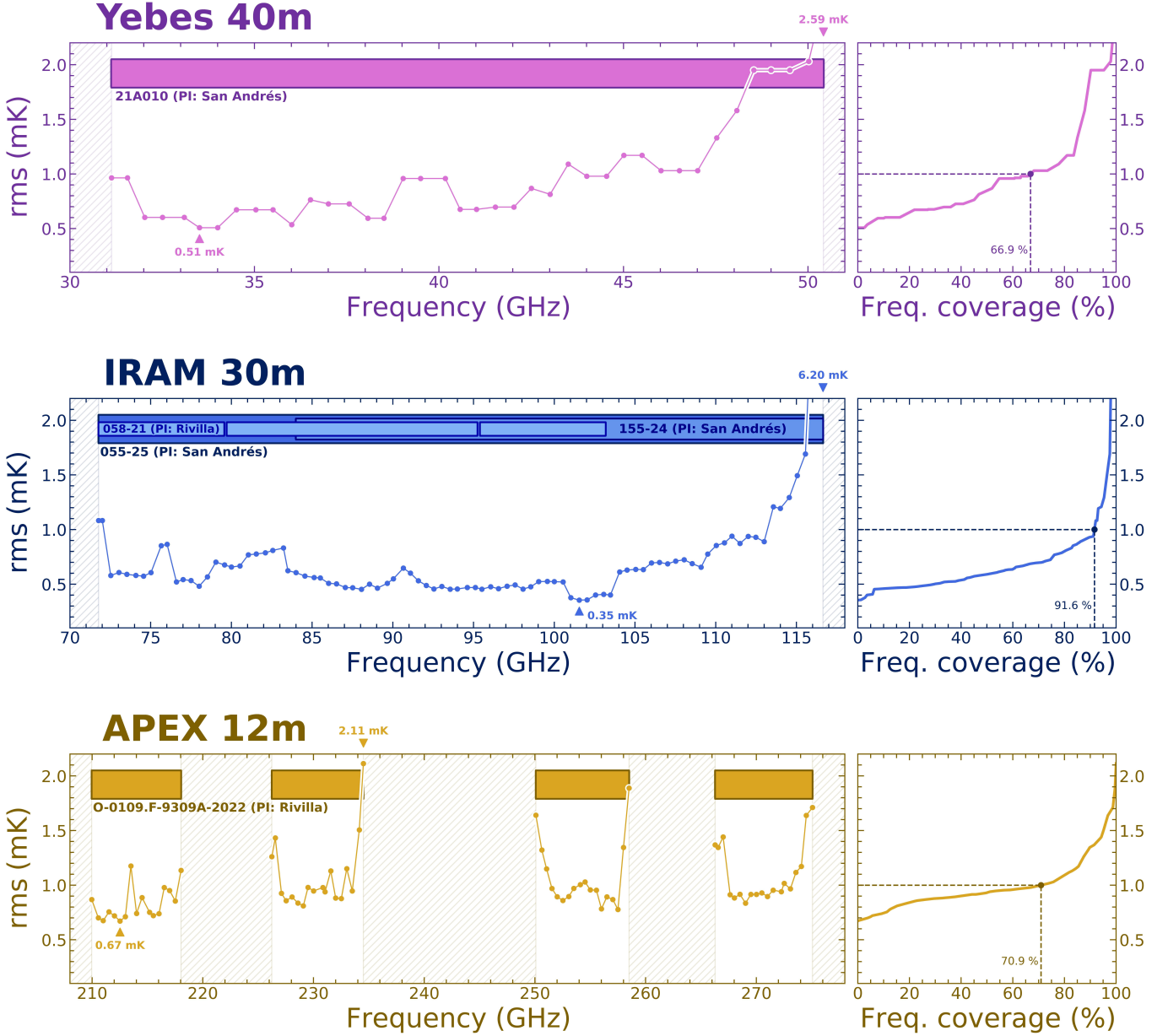}
    \caption{Spectral coverage and sensitivity (rms) achieved within the current molecular line survey of the G+0.633 molecular cloud, for the three single-dish telescopes used in this work. From top to bottom: Yebes 40m (magenta), IRAM 30m (blue), and APEX 12m (gold). The rms were estimated under a spectral resolution of 256\kHz for the Yebes 40m observations (translating into $\sim$1.54$-$2.56\kms in the covered range), 676\kHz for the IRAM 30m data ($\sim$1.74$-$2.82\kms) and 1\MHz ($\sim$1.1$-$1.5\kms) for the APEX survey.
    Left: measured rms across the observed frequency ranges, with triangles marking the minimum (upward triangle) and maximum (downward triangle) rms values in each dataset at their corresponding frequencies. The spectral coverage of the different observing programmes is shown by coloured superimposed boxes, including the project ID and PI. Right: achieved rms as a function of the cumulative spectral coverage, indicating the fraction of the observed bandwidth with each telescope reaching a sensitivity better than a given rms threshold. This representation highlights both the frequency-dependent behaviour of the noise and the statistical distribution of sensitivity across the surveyed bands for each telescope. The dashed lines indicate the fraction of spectral coverage within each particular telescope survey for which a sub-mK sensitivity (rms $\leq$ 1\mK) is achieved.}
    \label{fig:rms} 
\end{figure*}

Fig.~\ref{fig:rms} summarises the spectral coverage and the sensitivity (rms) achieved across the observed frequency ranges of the current G+0.633 molecular line survey, for the three radio telescopes used in this work (Yebes 40m, IRAM 30m, and APEX). The spectral windows of the different observational projects are also highlighted. %Together, these diagnostics provide a comprehensive overview of the survey coverage and the frequency-dependent behaviour of the noise across the observed spectral bands.

%Fig.~\ref{fig:rms} summarises the achieved rms sensitivity across the full frequency coverage of the G+0.633 spectral survey, for the three radio telescopes used in this work (Yebes 40m, IRAM 30m and APEX). This provides a global view of the noise properties and the effective sensitivity distribution as a function of frequency.

\clearpage

\section{Detection figures of cyanomethanimine isomers ($Z,E$-\ce{HNCHCN} and \ce{H2CNCN}), \ce{HOCH2CN}, $Z$-\ce{(CHOH)2}, \ce{C2H3NH2}, $sp$-$ap$-\ce{HOCOOH}, \ce{C2H5NCO}, \ce{(CH3)2CCH2}, $Z$-\ce{HNSO} and \ce{HOCS^+} towards G+0.633}
\label{appendix:molecules_additional_detection_figures}

\begin{figure*}[ht!]
    \centering
    \includegraphics[width=0.98\textwidth]{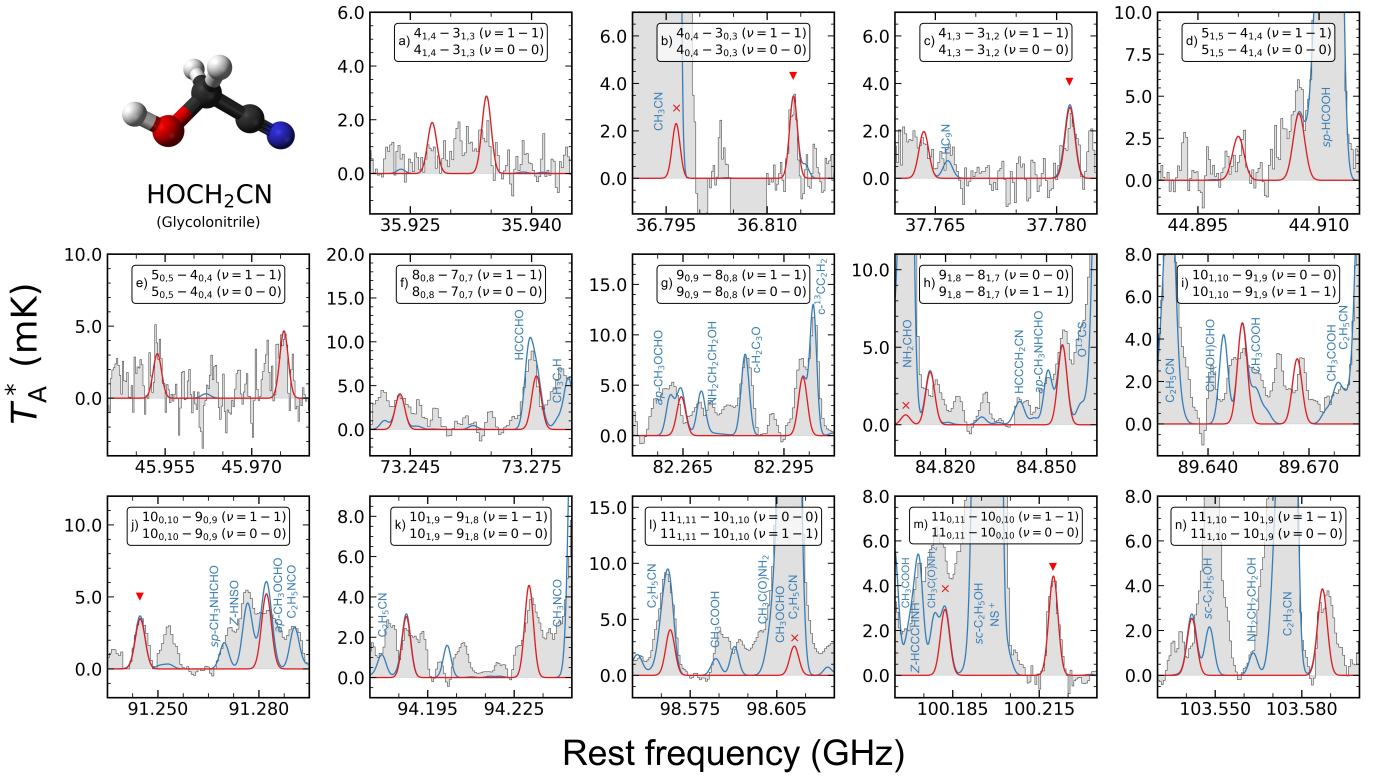}
    \caption{Same as Figs.~\ref{fig:urea_detection}$-$\ref{fig:CH3SCH3_detection}, but for \ce{HOCH2CN} (glycolonitrile). 
    %The spectroscopic details for all these transitions is given in Table~\ref{tab:HOCH2CN_rotational_spectroscopy} within Appendix~\ref{appendix:molecular_spectroscopy}. 
    Panel labels denote \ce{HOCH2CN} rotational transitions, labelled as $J_{K_\text{a}, K_\text{c}}$ and including their corresponding vibrational level ($\nu$). The four transitions within panels b), h), l) and m) marked with a $\times$ symbol were excluded in the fit due to significant blending, but shown here for consistency. The molecular structure of \ce{HOCH2CN} is depicted in the upper left corner, following the same colours palette as in the figures of the main text.} 
    \label{fig:HOCH2CN_detection} 
\end{figure*}

\begin{figure*}[ht!]
    \centering
    \includegraphics[width=0.98\textwidth]{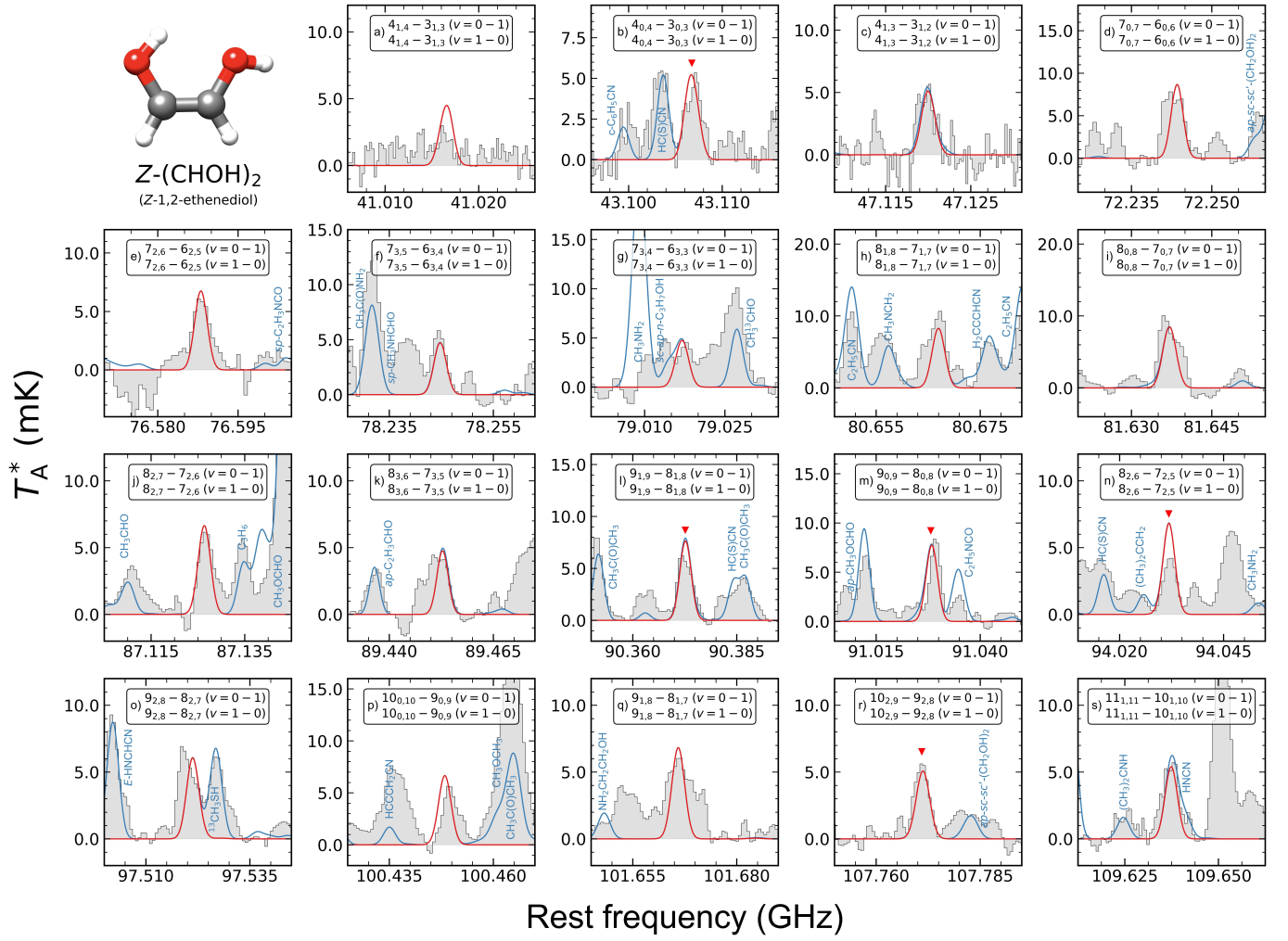}
    \caption{Same as Figs.~\ref{fig:urea_detection}$-$\ref{fig:CH3SCH3_detection}, but for $Z$-\ce{(CHOH)2} ($Z$-1,2-ethenediol). Panel labels denote $Z$-\ce{(CHOH)2} rotational transitions, labelled as $J_{K_\text{a}, K_\text{c}}$ and specifying the corresponding transition between the two inversion substates (\textit{v}) of the molecule.
    %Each of the observed lines correspond to the coalesced profiles emerging from the same rotational transition but between the two inversion substates (\textit{v}) of the molecule, namely the transitions $\text{\textit{v}}=0-1$ and $\text{\textit{v}}=1-0$.
    %The spectroscopic details for all these transitions is given in Table~\ref{tab:HOCH=CHOH_rotational_spectroscopy} within Appendix~\ref{appendix:molecular_spectroscopy}. Panel labels indicate the specific \ce{HOCH2CN} rotational transitions being plotted, which are labelled following the common notation for asymmetric tops ($J_{K_\text{a}, K_\text{c}}$) and include their corresponding vibrational level ($\nu$). 
    The molecular structure of $Z$-\ce{(CHOH)2} is depicted in the upper left corner, following the same colours palette as in the figures of the main text.} 
    \label{fig:HOCH=CHOH_detection} 
\end{figure*}

\begin{figure*}[ht!]
    \centering
    \includegraphics[width=0.98\textwidth]{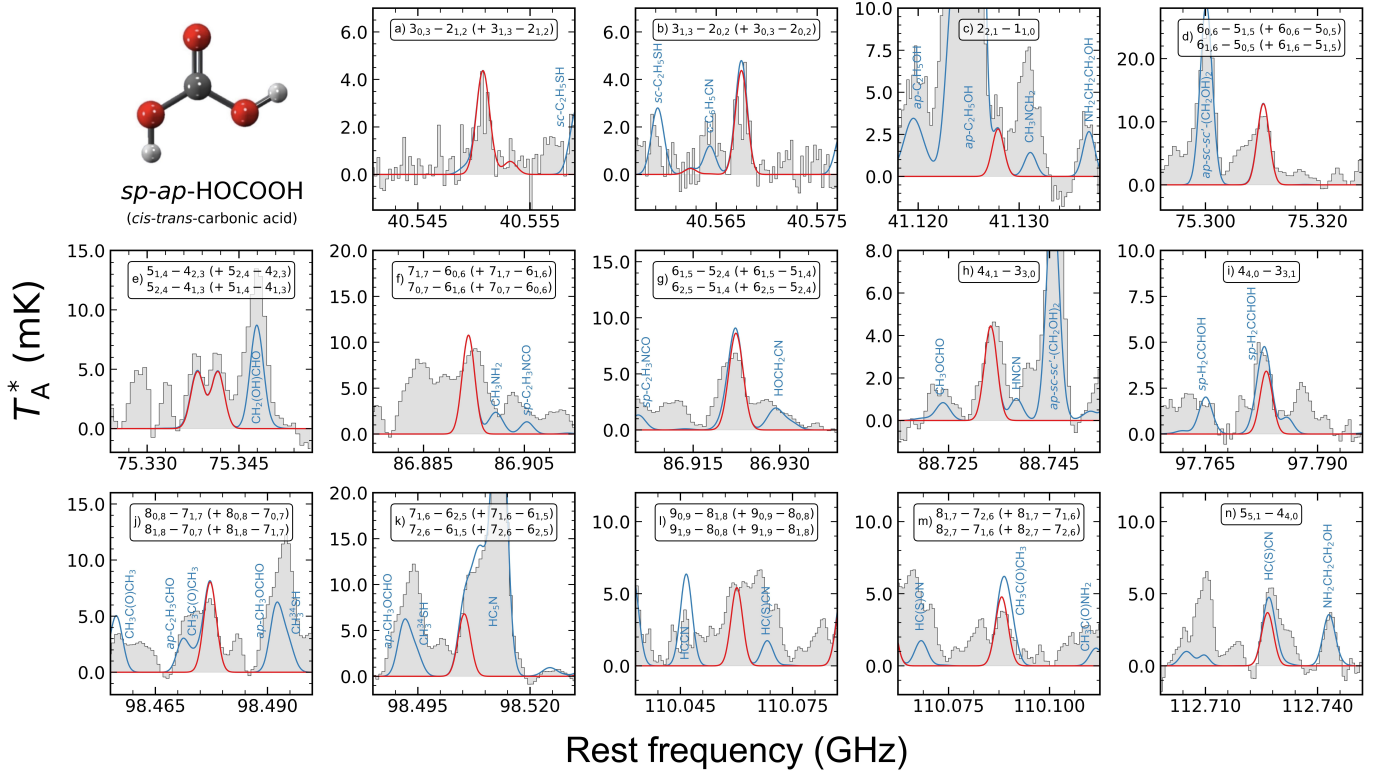}
    \caption{Same as Figs.~\ref{fig:urea_detection}$-$\ref{fig:CH3SCH3_detection}, but for $sp$-$ap$-\ce{HOCOOH} ($cis$-$trans$-carbonic acid). 
    %The spectroscopic details for all these transitions is given in Table~\ref{tab:cis-trans-HOCOOH_rotational_spectroscopy} within Appendix~\ref{appendix:molecular_spectroscopy}. 
    Transitions indicated in brackets are minor contributors to the spectral profile, but included in the fit. The molecular structure of $sp$-$ap$-\ce{HOCOOH} is depicted in the upper left corner, following the same colours palette as in the figures of the main text.} 
    \label{fig:cis-trans-HOCOOH_detection} 
\end{figure*}

\begin{figure*}[ht!]
    \centering
    \includegraphics[width=0.98\textwidth]{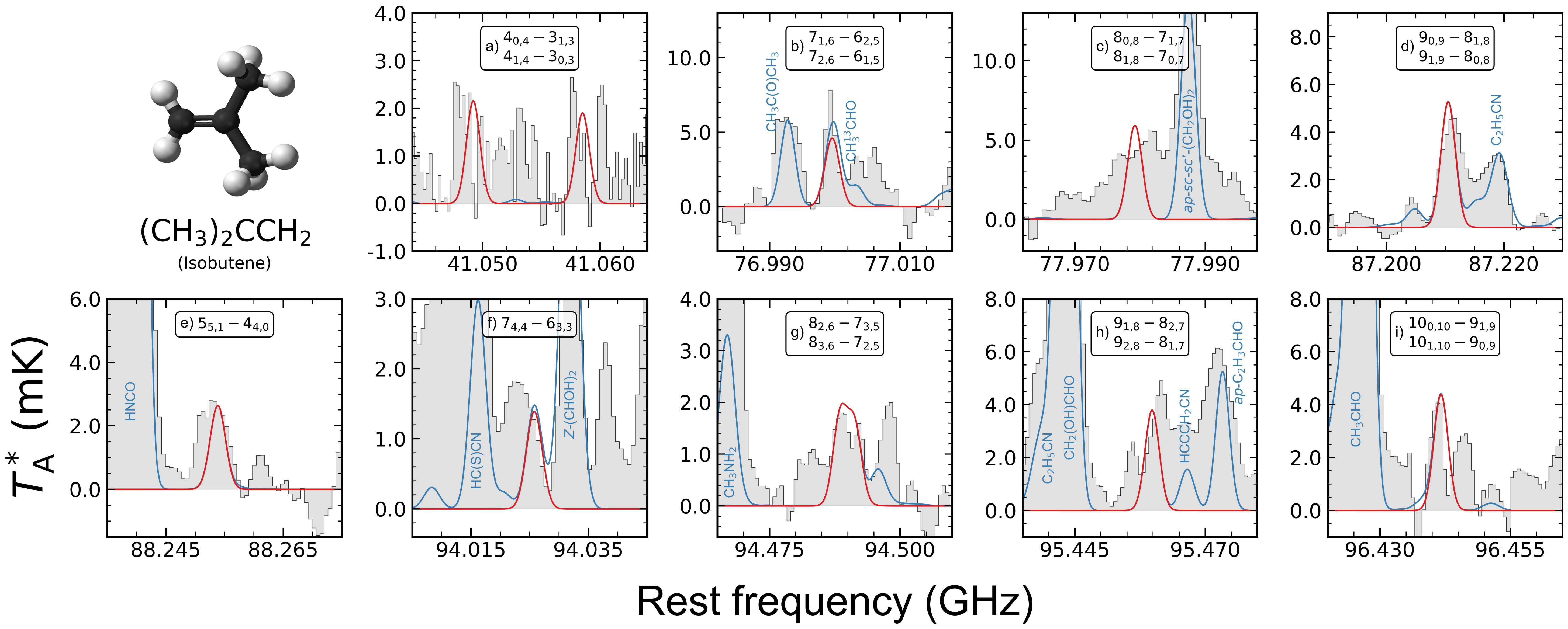}
    \caption{Same as Figs.~\ref{fig:urea_detection}$-$\ref{fig:CH3SCH3_detection}, but for \ce{(CH3)2CCH2} (isobutene). 
    %The spectroscopic details for all these transitions is given in Table~\ref{tab:Isobutene_rotational_spectroscopy} within Appendix~\ref{appendix:molecular_spectroscopy}. 
    The molecular structure of \ce{(CH3)2CCH2} is depicted in the upper left corner, following the same colours palette as in the figures of the main text.} 
    \label{fig:Isobutene_tentative} 
\end{figure*}

\begin{figure*}[ht!]
    \centering
    \includegraphics[width=0.98\textwidth]{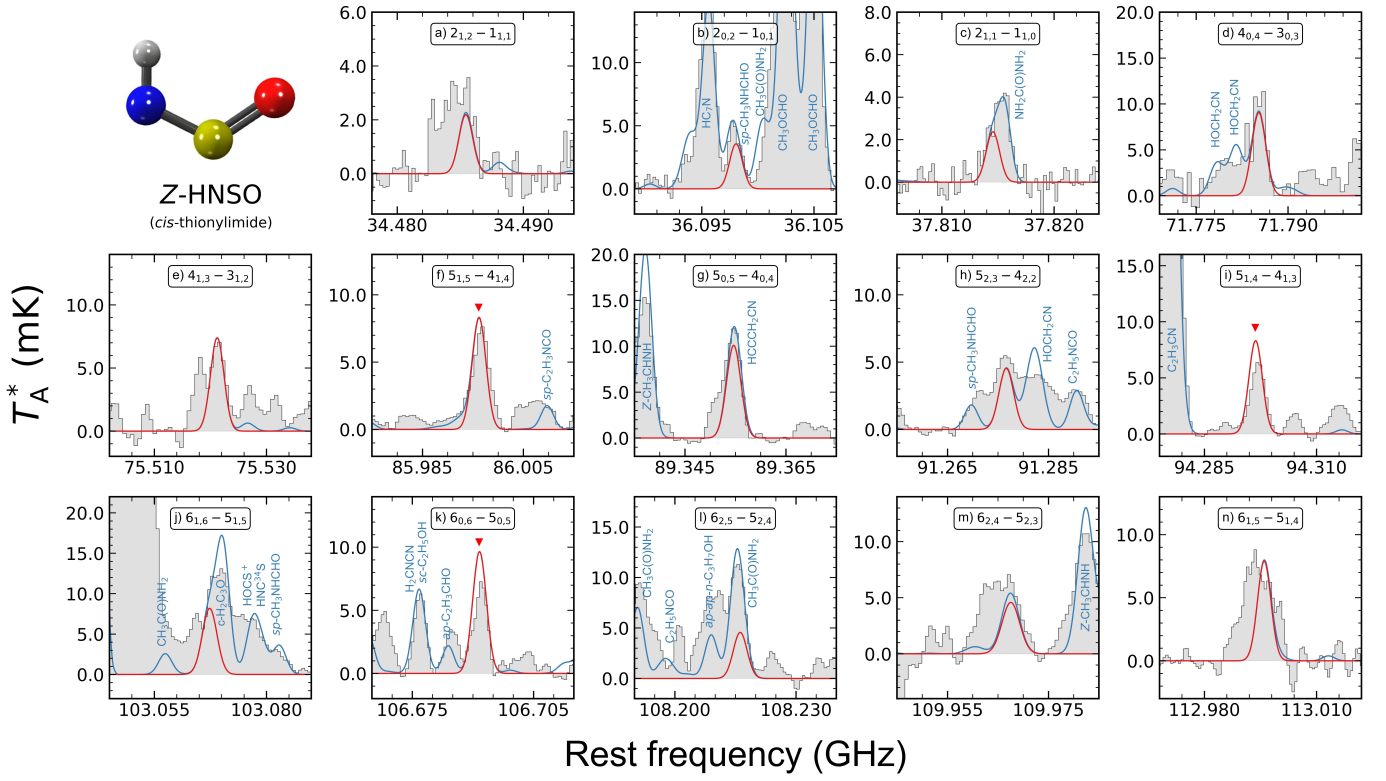}
    \caption{Same as Figs.~\ref{fig:urea_detection}$-$\ref{fig:CH3SCH3_detection}, but for $Z$-\ce{HNSO} ($cis$-thionylimide). 
    %The spectroscopic details for all these transitions is given in Table~\ref{tab:cis-HNSO_rotational_spectroscopy} within Appendix~\ref{appendix:molecular_spectroscopy}.
    The molecular structure of $Z$-\ce{HNSO} is depicted in the upper left corner, following the same colours palette as in the figures of the main text.} 
    \label{fig:cis-HNSO_detection} 
\end{figure*}

\begin{figure*}[ht!]
    \centering
    \includegraphics[width=0.98\textwidth]{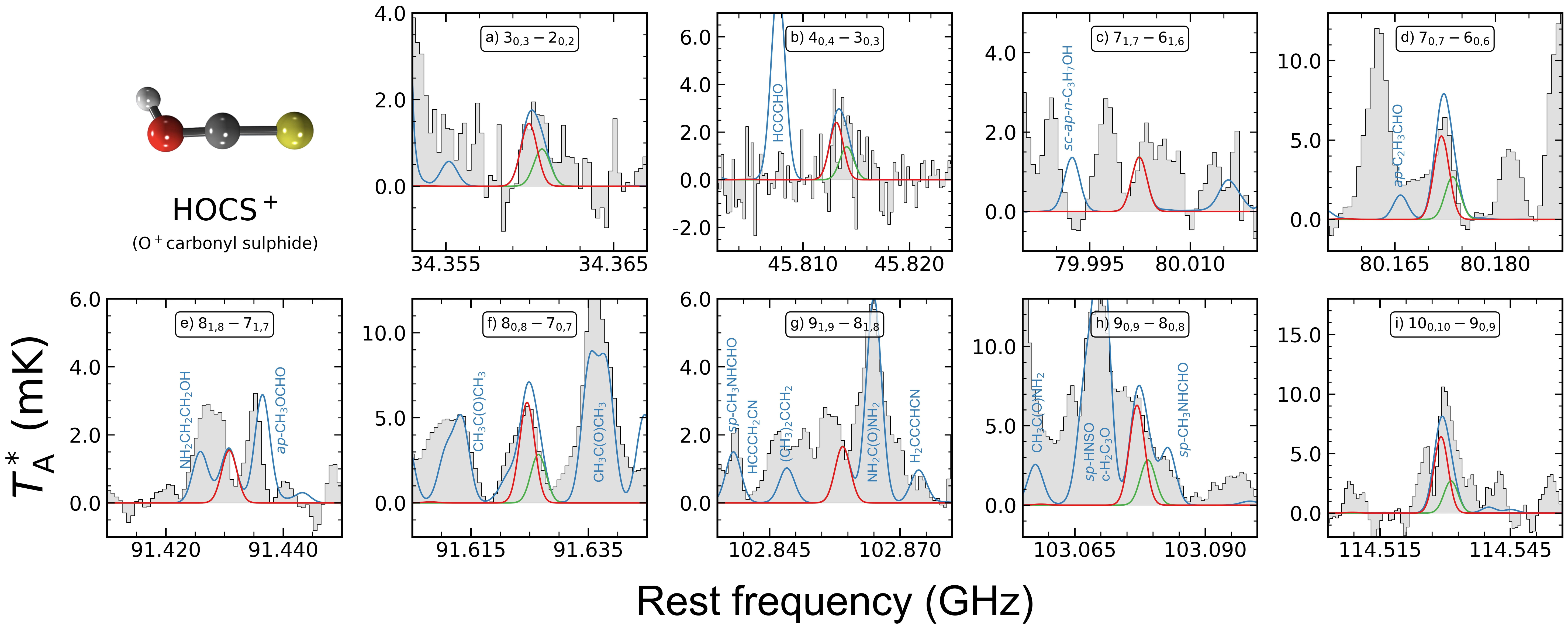}
    \caption{Same as Figs.~\ref{fig:urea_detection}$-$\ref{fig:CH3SCH3_detection}, but for \ce{HOCS^+} (\ce{O}-protonated carbonyl sulphide). The green solid lines in panels a), b), d), f), h) and i) represent the contribution from blending \ce{HNC^{34}S} emission affecting only the $K_{\text{a}} = 0$ transitions (see text). All remaining graphical elements are as in the main-text figures.
    %The spectroscopic details for all these transitions is given in Table~\ref{tab:cis-HNSO_rotational_spectroscopy} within Appendix~\ref{appendix:molecular_spectroscopy}.
    The molecular structure of \ce{HOCS^+} is depicted in the upper left corner.} 
    \label{fig:HOCS+_tentative} 
\end{figure*}

Figs.~\ref{fig:Z-HNCHCN_detection} to \ref{fig:HOCS+_tentative} present the detections of $Z,E$-\ce{HNCHCN}, \ce{H2CNCN}, \ce{HOCH2CN}, $Z$-\ce{(CHOH)2}, \ce{C2H3NH2}, $sp$-$ap$-\ce{HOCOOH}, \ce{C2H5NCO}, \ce{(CH3)2CCH2}, $Z$-\ce{HNSO} and \ce{HOCS^+} towards the G+0.633 molecular cloud, not included in the main text.

\begin{figure*}[ht!]
    \centering
    \includegraphics[width=0.98\textwidth]{Z-HNCHCN_NOLEGEND.jpeg}
    \caption{Same as Figs.~\ref{fig:urea_detection}$-$\ref{fig:CH3SCH3_detection}, but for $Z$-\ce{HNCHCN} ($Z$-$C$-cyanomethanimine). 
    %The spectroscopic details for all these transitions is given in Table~\ref{tab:Z,E-HNCHCN_H2CNCN_rotational_spectroscopy} within Appendix~\ref{appendix:molecular_spectroscopy}. 
    The molecular structure of $Z$-\ce{HNCHCN} is depicted in the upper left corner, following the same colours palette as in the figures of the main text.} 
    \label{fig:Z-HNCHCN_detection} 
\end{figure*}

\begin{figure*}[ht!]
    \centering
    \includegraphics[width=0.98\textwidth]{E-HNCHCN_NOLEGEND.jpeg}
    \caption{Same as Figs.~\ref{fig:urea_detection}$-$\ref{fig:CH3SCH3_detection}, but for $E$-\ce{HNCHCN} ($E$-$C$-cyanomethanimine). 
    %The spectroscopic details for all these transitions is given in Table~\ref{tab:Z,E-HNCHCN_H2CNCN_rotational_spectroscopy} within Appendix~\ref{appendix:molecular_spectroscopy}. 
    The molecular structure of $E$-\ce{HNCHCN} is depicted in the upper left corner, following the same colours palette as in the figures of the main text.} 
    \label{fig:E-HNCHCN_detection} 
\end{figure*}

\begin{figure*}[ht!]
    \centering
    \includegraphics[width=0.98\textwidth]{H2CNCN_NOLEGEND.jpeg}
    \caption{Same as Figs.~\ref{fig:urea_detection}$-$\ref{fig:CH3SCH3_detection}, but for \ce{H2CNCN} ($N$-cyanomethanimine). 
    %The spectroscopic details for all these transitions is given in Table~\ref{tab:Z,E-HNCHCN_H2CNCN_rotational_spectroscopy} within Appendix~\ref{appendix:molecular_spectroscopy}. 
    The molecular structure of \ce{H2CNCN} is depicted in the upper left corner, following the same colours palette as in the figures of the main text.} 
    \label{fig:H2CNCN_detection} 
\end{figure*}

\begin{figure*}[ht!]
    \centering
    \includegraphics[width=0.98\textwidth]{C2H3NH2_NOLEGEND.jpeg}
    \caption{Same as Figs.~\ref{fig:urea_detection}$-$\ref{fig:CH3SCH3_detection}, but for \ce{C2H3NH2} (vinylamine). 
    %The spectroscopic details for all these transitions is given in Table~\ref{tab:C2H3NH2_rotational_spectroscopy} within Appendix~\ref{appendix:molecular_spectroscopy}. 
    The molecular structure of \ce{C2H3NH2} is depicted in the upper left corner, following the same colours palette as in the figures of the main text.} 
    \label{fig:C2H3NH2_detection} 
\end{figure*}

\begin{figure*}[ht!]
    \centering
    \includegraphics[width=0.98\textwidth]{C2H5NCO_NOLEGEND.jpeg}
    \caption{Same as Figs.~\ref{fig:urea_detection}$-$\ref{fig:CH3SCH3_detection}, but for \ce{C2H5NCO} (ethyl isocyanate). 
    %The spectroscopic details for all these transitions is given in Table~\ref{tab:C2H5NCO_rotational_spectroscopy} within Appendix~\ref{appendix:molecular_spectroscopy}. 
    The molecular structure of \ce{C2H5NCO} is depicted in the upper left corner, following the same colours palette as in the figures of the main text.} 
    \label{fig:C2H5NCO_detection} 
\end{figure*}

\end{appendix}

\end{document}